\documentclass[aps,pra,showpacs,twoside,twocolumn,10pt]{revtex4-2}
\usepackage[colorlinks=true, citecolor=red, urlcolor=blue ]{hyperref}
\usepackage{epsfig,newlfont,amssymb,amsfonts,amsmath,bm,subfigure,palatino,mathtools,amsthm,braket,times,soul,enumitem,color,xcolor}
\usepackage[normalem]{ulem}
\newcommand{\stkout}[1]{\ifmmode\text{\sout{\ensuremath{#1}}}\else\sout{#1}\fi}
\usepackage{enumitem}

\colorlet{saveblue}{blue}

\begin{document}
\title{All multipartite entanglements are quantum coherences in locally distinguishable bases}

\author{Ahana Ghoshal$^{1,2}$, Swati Choudhary$^{1,3}$, and Ujjwal Sen$^1$}
\affiliation{$^1$Harish-Chandra Research Institute, A CI of Homi Bhabha National Institute, Chhatnag Road, Jhunsi, Prayagraj 211 019, India\\
$^{2}$ Naturwissenschaftlich-Technische Fakult\"{a}t, Universit\"{a}t Siegen, Walter-Flex-Stra\ss e 3, 57068 Siegen, Germany\\
$^3$International Institute of Information Technology Hyderabad, Prof. CR Rao Road, Gachibowli, Hyderabad 500 032, Telangana, India}

\begin{abstract}
We find that the $m$-separability and $k$-partite entanglement of a multipartite quantum system is correlated with quantum coherence of the same with respect to complete orthonormal bases, distinguishable under local operations and classical communication in certain partitions. In particular, we show that the geometric measure of $m$-inseparable entanglement of a multipartite quantum state is equal to the square of minimum fidelity-based quantum coherence of the state with respect to complete orthonormal bases, which are locally distinguishable in a partition into $m$ parties.
\end{abstract}
\maketitle

\section{Introduction}
\label{sec:intro}
Entanglement~\cite{ent-review,GUHNE20091,Das2019} and quantum coherence~\cite{Aberg2006,coh,Winter2016,Streltsov2017} play central roles in quantum information protocols and have become the backbone in the development of quantum resource theories in the last few decades. Entanglement acts as a key resource in several quantum communication protocols including quantum dense coding~\cite{supcod,Guo2019}, quantum teleportation~\cite{tele,Pirandola2015,Liu_2020}, and entanglement-based quantum cryptography~\cite{qcryp,Gisin2002,Pirandola20,Portmann2021}. Coherence of a quantum state, which, like entanglement, is an outcome of the superposition principle of quantum mechanics, allows one to understand phenomena like interference patterns in interferometers and plays an important role in e.g. the study of lasers~\cite{Lukin1998}, quantum-enhanced metrology~\cite{Pires2018,Castellini2019,Zhang2019,Sun2022}, quantum algorithms~\cite{Hillery2016,Anand2016,Shi2017,Liu2019,Shubhalakshmi2020}, quantum state
discrimination~\cite{Xiong_2018,Kim2018}, quantum thermodynamics~\cite{Lostaglio2015,thermo,Misra2016,Lostaglio_2017}, and possibly in biology~\cite{Huelga2013}.
However, the two quintessential quantum resources, quantum entanglement and quantum coherence, although interconnected in many ways, possess significant differences. In particular, entanglement is a basis-independent quantity, whereas quantum coherence is naturally dependent on the choice of basis. Studies on the inter-relations between the two resources include Refs.~\cite{Oppenheim2003,Asboth2005,Yao2015,Streltsov2015,Xi2015,Chitambar2016,Killoran2016,Streltsov2016,Qi_2017,Zhu2017,Chin2017,Zhu2018,Egloff2018,Streltsov_2018,Kraemer2021,relation_ent-coh,Kim2022,Liu2022,Bhattacharyya2021,Zhang2024}.\par
The characterisation and detection of multipartite quantum entanglement are an interesting and useful area of research in quantum information~\cite{ent-review,GUHNE20091,Das2019,Laskowski2010,G_hne_2010,Huber2010,Gabriel2010,Ananth2015,Das2017}. Multipartite quantum states can be categorised into different groups on the basis of their separability and entanglement properties. The fully separable and genuine multipartite entangled states~\cite{G_hne_2010,Huber2010} are two important groups. The genuine $n$-partite entanglement of an $n$-partite quantum state can be identified by various approaches~\cite{G_hne_2010,Huber2010,Gao2010,Gao2011,Wu2012,Chen2012}, like spin-squeezing inequalities~\cite{Vitagliano2011}, Bell-type inequalities~\cite{Seevinck2002}, state extensions~\cite{Doherty2002,Doherty2004,Doherty2005}, and covariance matrices~\cite{Gittsovich2010}. For multipartite quantum systems, there exist some states which are neither fully separable nor genuinely entangled. These states can be categorised as ``$m$-separable" ones on the basis of their separability properties and as $k$-partite entangled ones on the basis of the maximum number of entangled parties
present in at least one cluster of the separable partitions. 
For research on $m$-separability and $k$-partite entanglement, see e.g.~\cite{Huber2010,Gabriel2010,Hyllus2012,Toth2012,Hong2012,Huber2013,Gao_2013,Gao2014,Klockl2015,Hong2015,Ananth2015,Gessner2016,Das2017,Hong2016,HONG2021}. A new property called $r$-stretchability was introduced to detect multipartite entanglement, unifying the advantages of $m$-separability and $k$-partite entanglement~\cite{Szalay2019,Toth2020}. Furthermore, Ref.~\cite{Ren2021} introduced an approach to characterise metrologically useful multipartite entanglement through the use of Young's diagrams to represent different partitions.

A complete orthonormal basis of (pure) quantum states of a system of two or more subsystems, distinguishable under local quantum operations and classical communication (LOCC)~\cite{LOCC1,LOCC2,Rains1998,LOCC3}, is called a locally distinguishable basis. 
An orthonormal basis of a composite quantum system is locally distinguishable by LOCC only if all basis states are product states across the relevant subsystems~\cite{Horodecki2003}. 
Local distinguishability of orthogonal states of multiparty systems is rather unconnected to the entanglement of the constituent systems~\cite{Bennett1999,Walgate2000,Horodecki2003}. 
However, Ref.~\cite{relation_ent-coh} (see also~\cite{Zhu2017}), found that entanglement can be qualitatively as well as quantitatively defined via quantum coherence with respect to all locally distinguishable complete orthonormal bases. The results in~\cite{relation_ent-coh} were mostly for bipartite systems. Entanglement however can be and has been generalised to the multipartite domain. Multiparty entangled states present far more challenges in attempts to characterise them, while having much more potential utility than the bipartite variety.
\par
In this paper, we find a relation between the $m$-separability of a multipartite quantum system and its quantum coherence with respect to complete orthonormal bases, locally distinguishable in certain partitions.
We also show that a similar relation holds for $k$-partite entanglement and quantum coherence with respect to the same for a multipartite quantum system. 
We subsequently quantify the connection between $m$-separability and quantum coherence by showing that for a pure multiparty quantum state, the square of the minimum of fidelity-based quantum coherence~\cite{rel_entropy,vedral_ent} with respect to the optimal complete orthonormal basis, locally distinguishable in partitions into $m$ parties, turns out to be equal to the geometric measure of entanglement~\cite{GM,Blasone2008,Cianciaruso2016,Das2017}, which is a measure of $m$-inseparability of a multipartite entangled state. We extend this quantification for multipartite mixed states also.\par
The rest of this paper is presented as follows. In Sec.~\ref{sec:pre}, we provide the basic ideas of $m$-separable states and $k$-partite entangled states and  the measures of quantum entanglement and quantum coherence considered for obtaining the results of our paper. Section~\ref{sec:res} contains the theorems and the corresponding proofs, which connect the quantum entanglement and quantum coherence of a multipartite quantum system. A summary is given in Sec.~\ref{sec:con}.
 
 \section{Underlying tools}
 \label{sec:pre} 
For a multipartite quantum system, the definition of an entangled quantum state is layered. If an $n$-partite pure quantum state ($n>2$) is separable in all its partitions, the state is called an $n$-separable or a fully separable state. On the other hand, if the state is entangled in at least one bipartite splitting, then it is an entangled one.
These ``entangled" states can however be categorised into different classes in various ways. 
The definitions of $m$-separability and $k$-partite entanglement of an $n$-partite quantum state, and a brief discussion of the basic concepts of the two, are briefly presented now.
\\


\noindent \textit{$m$-separability}: An \textit{n-partite} pure quantum state is called $m$-separable if it can be written as a product of pure states of a maximum of $m$  subsystems, viz.
     \begin{equation}
     \ket{\psi_{\textit{m-\text{sep}}}}=\ket{\psi_{1}}\otimes\ket{\psi_{2}}\otimes\ldots\otimes\ket{\psi_{m}},
     \label{Eq:1}
     \end{equation}
where $m$ can be any integer  $\le n$. \\

We will refer to a state in any of the subsystems as a ``substate''. 
An $n$-partite pure quantum state is called fully separable if and only if $m=n$, and it is genuinely multipartite entangled if and only if the state is not biseparable, i.e., $m=1$.
A quantum state $\rho_{\textit{$m$-\text{sep}}}$, possibly mixed, is $m$-separable if 
(i) it has at least one pure-state decomposition, in which there is at least one $m$-separable pure state and no other state with separability lower
than $m$ in that decomposition, and 
(ii) there is no other pure-state decomposition with all pure states that are $\tilde{m}$-separable with $\tilde{m}\ge m+1$. Therefore, there exists a probabilistic decomposition,
     \begin{equation}
     \label{m-sep_mixed}
     \rho_{m\textit{-\text{sep}}}=\sum_{i}p_{i}\ket{\psi^{i}}\bra{\psi^{i}},
     \end{equation}
where for at least one $i$, $\ket{\psi^i}$ will be $\ket{\psi_{\textit{m-sep}}}$ and, for the other $i$'s,  $\ket{\psi^i}$ will be $\ket{\psi_{m^{\prime}\textit{-sep}}}$, where $m^{\prime}\ge m$. Furthermore, there does not exist a decomposition for which every element is $\tilde{m}$-separable  with $\tilde{m} \ge m+1$. If there exist $m^{\prime}$'s which are equal to $m$, then the individual $m$-separable pure states composing the $n$-partite and $m$-separable mixed state can be $m$-separable in different partitions. Hence, in general, $m$-separable mixed states are not separable with respect to any specific partition, which makes $m$-separability rather difficult to detect. An $n$-partite pure quantum state being $m$-separable implies that the state will exhibit some entanglement between some parts of the system, provided $n>m$. The degree of entanglement of an $n$-partite entangled state can be described by $k$-partite entanglement, which is defined below.
\\

\noindent\textit{$k$-partite entanglement}: Any $n$-partite pure quantum state
is called $k$-partite entangled 
if there exists at least one genuinely $k$-partite entangled substate and no other pure substate has genuinely $\tilde{k}$-party entanglement for $\tilde{k} \ge k+1$.\\

There can be $(k-1)$-partite entangled substates present in such states, but the maximum degree of entanglement is $k$. Like in the $m$-separability case, we can define $k$-partite entanglement for mixed states also. An $n$-partite 
quantum state, 
possibly mixed, is called $k$-partite entangled when (i) there exists at least one decomposition into pure states, in which at least one pure state is genuinely $k$-partite entangled, (ii) no other state exhibits $\ge (k+1)$-partite entanglement, and (iii) there is no other pure-state decomposition with all pure states that is genuinely $\tilde{k}$-party entangled with $\tilde{k}\le k-1$.
Various techniques~\cite{GUHNE20091,Friis2019} enable the determination of experimentally observable bounds for $k$ and $m$, while the difference $k-(m+1)$ serves as a valuable resource in quantum metrology~\cite{Ren2021}. A structured analysis of multipartite system partitions further revealed a fundamental duality between $k$ and
$m+1$~\cite{Szalay2019}. However, two different $n$-partite and $m$-separable quantum states with the same values of $n$ and $m$ can be $k$-partite entangled for two different values of $k$, and the parallel statement is also true when we interchange $k$-partite entanglement and $m$-separability. This fact is clearly visible from the following example. Let us consider two pure 10-partite quantum states,
\begin{enumerate}[label=(\roman*)]
\item   $\ket{GHZ}\ket{GHZ}\ket{GHZ}\ket{0}$
\item $\ket{GHZ}\ket{\psi^{-}}\ket{\psi^{-}}\ket{\psi^{-}}\ket{0}$,
\end{enumerate}
where $\ket{GHZ}$ is the Greenberger-Horne-Zeilinger state (GHZ state)~\cite{Greenberger1989,Mermin1990} for three-qubits and $\ket{\psi^{-}}$ is one of the Bell states:
\begin{eqnarray}
    &&\ket{GHZ}=\frac{1}{\sqrt{2}}(\ket{000}+\ket{111}),\nonumber\\
    &&\ket{\psi^{-}}=\frac{1}{\sqrt{2}}(\ket{01}-\ket{10}).
\end{eqnarray}
Here $\ket{0}$ and $\ket{1}$ indicate, respectively, the excited- and ground-state eigenvectors of the Pauli-$z$ operator. Both states, (i) and (ii), are three-partite entangled as the maximum number of entangled qubits is three in both the cases, but (i) is a four-separable state while (ii) is a five-separable one. 
Similarly, there exist multipartite states which can be written in the same separable structure but exhibit different degrees of entanglement.\par
The complexity of the structure of $m$-separable and $k$-partite entangled states makes their detection for an $n$-partite system with $n>2$ rather challenging. There are several multipartite entanglement measures known in the literature. In this paper, we concentrate on the geometric measures ($GM$s) of entanglement~\cite{GM,Blasone2008,Cianciaruso2016,Das2017,GMsymm,zhu2010,Zhang2016}.
The $GM$s 
are distance-based measures of entanglement for a multipartite quantum state and are defined by keeping an appropriate class of separable states as the reference states. 
For clarity of notation, we use $GM_m$ for the geometric measure of entanglement
in the further discussions to specify the $m$-separable reference states. We now provide definitions of and a brief discussion of a few concepts that will be useful in this paper.
\\

\noindent \textit{Geometric measure of entanglement}: This is a measure of $m$-inseparability of a multipartite entangled state that is quantified by the distance of a given entangled state from the nearest $m$-separable one. For a multipartite pure state $\ket{\psi}$, the geometric measure of entanglement is defined as
     \begin{equation}
     \label{GM_pure}
       GM_m(\ket{\psi})=\min_{\ket{\phi}}(1-|\braket{\phi|\psi}|^2),  
     \end{equation}
where $\ket{\phi}$ belongs to the set of $m$-separable pure states. \\

If this $\ket{\phi}$ belongs to the set of biseparable pure states, then the multipartite entanglement measure $GM_2$ will be a measure of genuine multipartite entanglement, the generalised geometric measure $(GGM)$~\cite{Sen(De)2010,Sen2010,Biswas2014,GGM}. For a multipartite mixed state $\rho$, the geometric measure of entanglement can be defined using the convex roof approach~\cite{LOCC1,GGM}. Thus,
\begin{equation}
\label{GM_mixed}
    GM_m(\rho)=\min_{\{p_i,\ket{\psi^i}\}}\sum_i p_i GM_m(\ket{\psi^i}),
\end{equation}
where the minimisation is over all  pure state decompositions of $\rho=\sum_i p_i\ket{\psi^{i}}\bra{\psi^{i}}$.
\\

We now move over to the next concept that we will need in this paper, viz. that of multi-orthogonal product bases. 
\\
\\
\noindent\textit{Multi-orthonormal product basis}: This is a basis in the Hilbert space $C_G \equiv \mathbb{C}^{d_{1}}\otimes\mathbb{C}^{d_{2}}\otimes\ldots\otimes \mathbb{C}^{d_{n}}$, whose elements can be written in the product form of the constituent substates of $\mathbb{C}^{d_i}$ where $i$ runs from $1$ to $n$, and the local pure states form orthonormal bases in $\mathbb{C}^{d_i}$. Therefore, a multi-orthonormal (fully separable) product basis in $C_G$ has the form
\begin{equation}
    \{\ket{\psi_{j_1}^1}\otimes\ket{\psi_{j_2}^2}\otimes\ldots\otimes\ket{\psi_{j_n}^n}\}_{j_1,j_2,\ldots,j_n},
\end{equation}
where $j_1$ runs from $1$ to $d_1$, $j_2$ runs from $1$ to $d_2$ and so on and $\{\ket{\psi_{j_i}^i}\}_{j_i}$ forms an orthonormal basis in $\mathbb{C}^{d_i}$ for $i=1,2,\ldots,n$.
Similarly, the structure of an $m$-separable multi-orthonormal product basis in the Hilbert space $C_G$ with a corresponding partition being as in \(\mathbb{C}^{\tilde{d}_{1}}\otimes\mathbb{C}^{\tilde{d}_{2}}\otimes \ldots \otimes\mathbb{C}^{\tilde{d}_{m}}\), with \(\prod_{i=1}^n d_i = \prod_{i=1}^m \tilde{d}_i\), takes the form
\begin{equation}
    \{\ket{\psi_{\tilde{j}_1}^1}\otimes\ket{\psi_{\tilde{j}_2}^2}\otimes\ldots\otimes\ket{\psi_{\tilde{j}_m}^m}\}_{\tilde{j}_1,\tilde{j}_2,\ldots,\tilde{j}_m}.
    \label{Eq:m_sep_basis}
\end{equation}
Here $\tilde{j}_1$ runs from $1$ to $\tilde{d}_1$, $\tilde{j}_2$ runs from $1$ to $\tilde{d}_2$ and so on.
It is important to note that it is always possible to construct an \(m\)-separable multi-orthonormal product basis from any $n$-partite $m$-separable pure state. Here we outline a general construction. Let us
consider a given $m$-separable state
\begin{equation}
\ket{\Psi} = \ket{\Psi_1} \otimes \ket{\Psi_2} \otimes \dots \otimes \ket{\Psi_m},
\end{equation}
where each $|\Psi_i\rangle$ belongs to the Hilbert space $\mathbb{C}^{\tilde{d}_{i}}$ of the corresponding subsystem in the given partition. For each subsystem $i$, we now begin with $|\Psi_i\rangle$ as one of the basis elements and extend it to form an orthonormal basis $\{\ket{e_{\tilde{j}_i}^i}\}_{\tilde{j}_1=1}^{\tilde{d}_i}$ in $\mathbb{C}^{\tilde{d}_{i}}$. The global multi-orthonormal product basis for the total Hilbert space is then formed by taking all possible tensor products of these local bases, as in Eq.~(\ref{Eq:m_sep_basis}). 
Since each subsystem has an orthonormal basis, this set forms a complete orthonormal basis for the entire space. 
Thus, an $m$-separable multi-orthonormal 
product basis containing the given $m$-separable state $|\Psi\rangle$ can be constructed, and it is locally distinguishable since it is formed from the tensor products of orthonormal bases of the subsystems.

Note that for a mixed state, it is generally not possible to find a single product basis that simultaneously contains all product components of the mixture, since the pure states in a convex decomposition of a separable mixed state need not be mutually orthogonal or share a common set of local basis vectors. This, however, does not affect our analysis. When discussing the relation between $m$-separability of a multipartite quantum system and its quantum coherence with respect to complete orthonormal bases that are locally distinguishable under certain partitions, we adopt the convex-roof definition of coherence for mixed states. This definition considers all pure-state decompositions of the mixed state. For each pure product component in such a decomposition, one can always extend it to a complete multi-orthonormal product basis that is LOCC-distinguishable.
\\

We now move over to the discussion of another resource, viz. quantum coherence. Unlike entanglement, quantum coherence is a basis-dependent quantity. It is always defined with respect to some fixed basis, and changing the basis results in a change in coherence of a system state. The basic idea and the measure of quantum coherence used in this paper are presented below.\\

\noindent\textit{Quantum coherence}: A pure quantum state $\ket{\psi}$ of a physical system with Hilbert space $\mathbb{C}^d$ is claimed to be quantum coherent with respect to a complete orthonormal basis of $\mathbb{C}^d$ if the density matrix $\rho=\ket{\psi}\bra{\psi}$ is not diagonal when written in that basis~\cite{Aberg2006,coh,Winter2016,Streltsov2017}.\\

For a mixed state, the concept of quantum coherence can be defined by using the concept of convex roof~\cite{LOCC1,GGM}. A mixed quantum state $\rho$, on the Hilbert space $\mathbb{C}^d$, is said to be quantum coherent with respect to a class of complete orthonormal bases $\{B\}$ of 
  $\mathbb{C}^d$, if it can not be written as a convex sum of pure states of the same Hilbert space, with zero minimal quantum coherence when optimised over such bases~\cite{relation_ent-coh}. Therefore, for a quantum state $\rho$ in $\mathbb{C}^d$, quantum coherence with respect to the class $\{B\}$ of bases in $\mathbb{C}^d$ is given by
  \begin{equation}
  \label{rel_entro_coher_mixed}
     C_{\{B\}}(\rho) = \min \sum_{i} p_{i} \min_{B\in{\{B\}}} C_{B}(\ket{\psi^i}), 
  \end{equation} 
   where the outer minimisation is over all decompositions of $\rho$ into $\sum_{i} p_i \ket{\psi^i}\bra{\psi^i}$ and $C_{B}(\ket{\psi^i})$ is any measure of quantum coherence for pure states $\ket{\psi^i}$. In this paper, the measure is taken to be the fidelity-based coherence measure, discussed below.\\
   
\noindent \textit{Fidelity-based quantum coherence measure}: For a pure quantum state $\ket{\psi}$, of the Hilbert space $\mathbb{C}^d$, the fidelity-based quantum coherence with respect to a complete orthonormal basis $B$ can be defined as
     \begin{equation}
    \label{fid_coher}
    C_{B}^{F}(\ket{\psi})=\min_{{\ket{\tilde{\phi}}}\in I }\sqrt{1-F(\ket{\psi},\ket{\tilde{\phi}})},
    \end{equation} 
    where $I$ is the set of all incoherent pure states corresponding to basis $B$, i.e., the elements of $B$, and $F(\ket{\psi},\ket{\tilde{\phi}})=|\braket{\tilde{\phi}|\psi}|^2$. See e.g.~\cite{fid}.
    The  convex roof construction can lead us to obtain the measure for mixed state inputs, and therein 
    we have,
     \begin{equation}
    \label{rel_ent_coher}
    C_{B}^{F}(\rho)=\min_{\{p_i,\ket{\psi^i}\}}\sum_i{p_i  C_{B}^{F}(\ket{\psi^i}) }.
    \end{equation} 
    The minimisation is over all the pure state decompositions of $\rho=\sum_i p_i\ket{\psi^{i}}\bra{\psi^{i}}$.
\section{Connection of $m$-separability and $k$-partite entanglement with quantum coherence}    
\label{sec:res}
It was shown in~\cite{relation_ent-coh} (see also~\cite{Zhu2017}) that quantum coherence in locally distinguishable bases of bipartite systems can be connected to entanglement, both qualitatively and quantitatively. Multipartite systems however are known to possess a far richer structure, in comparison to bipartite ones, in terms of entanglement~\cite{parisio2017,kraus2018,multi}  as well as local distinguishability~\cite{multiloc}  and some connections between entanglement and quantum coherence in locally distinguishable bases were already alluded to in~\cite{relation_ent-coh}. 
We provide here a complete characterisation of this connection with respect to the entire hierarchy of $m$-separable and $k$-partite entangled states.\\
\\
\noindent\textit{Theorem 1}: \textit{An $n$-partite pure quantum state of the Hilbert space $\mathbb{C}^{d_{1}}\otimes\mathbb{C}^{d_{2}}\otimes\ldots \mathbb{C}^{d_{n}}$ is $m$-separable if and only if (i) it is quantum coherent with respect to all complete orthonormal bases that are locally distinguishable in an arbitrary partition into $m+1$ parties and (ii) it is incoherent with respect to at least one complete orthonormal basis locally distinguishable in at least one partition into $m$ parties.}\par
\textit{Proof:} \(\implies\) part. Let $\ket{\psi} \in \mathbb{C}^{d_{1}}\otimes\mathbb{C}^{d_{2}}\otimes\ldots \mathbb{C}^{d_{n}}$ be an $m$-separable state. Therefore, $\ket{\psi}=\ket{\psi_1}\otimes \ket{\psi_2}\ldots\otimes\ket{\psi_m}$, with $\ket{\psi_i}\in \mathbb{C}^{\overline{d}_k}$, where $\prod_{k=1}^m \overline{d}_k=\prod_{i=1}^n d_i$. Also, any $\ket{\psi_k}$ is either a single-party state or is genuinely multiparty entangled in $\mathbb{C}^{\overline{d}_k}$. Let $\{B_L^{m+1}\}$ be the set of all complete orthonormal bases 
locally distinguishable in an arbitrary partition of the $n$ parties into $m+1$ clusters. 
For a given element of $\{B_L^{m+1}\}$, let that partition be $A_1:A_2:\ldots:A_{m+1}$. Then $\ket{\psi}$ is quantum coherent with respect to that element of $\{B_L^{m+1}\}$~\cite{Horodecki2003,relation_ent-coh}. So, $\ket{\psi}$ is quantum coherent with respect to all the elements of $\{B_L^{m+1}\}$.

On the contrary, if we consider the set of all complete orthonormal bases, locally distinguishable in at least one partition into $m$ parties, $\ket{\psi}$
must be incoherent with respect to at least one such basis. The reason is the following. With the $m$-separable state $\ket{\psi}$, separable in the partition, say $A_1:A_2:\ldots:A_m$, 
we can always construct an $m$-separable multi-orthonormal product basis in $\mathbb{C}^{d_{1}}\otimes\mathbb{C}^{d_{2}}\otimes\ldots \mathbb{C}^{d_{n}}$, as discussed in Sec.~\ref{sec:pre}, which can be distinguished by LOCC in the partition $A_1:A_2:\ldots:A_m$, because it consists of tensor products of orthonormal bases from each subsystem. 
As $\ket{\psi}$ is an element of this basis, it is naturally incoherent with respect to the basis. Note that, regarding the construction of an $m$-separable, multi-orthonormal, locally distinguishable basis containing a given $m$-separable state, we do not assume that all $m$-separable multi-orthonormal product bases are locally distinguishable. Rather, we assert that at least one such basis can always be constructed from a given $m$-separable state. As we mentioned earlier, this is achieved by extending each local component of the given $m$-separable state to a full orthonormal basis in its respective subsystem. The tensor product of these local bases forms a global $m$-separable basis that is manifestly locally distinguishable: each party can measure in its local basis and communicate outcomes classically, ensuring perfect discrimination of the basis states. This construction guarantees the existence of a locally distinguishable basis containing the given $m$-separable state, without contradicting known counterexamples such as the basis that leads to ``quantum nonlocality without entanglement'', which correspond to special product bases that are not constructed in this way~\cite{Bennett1999}. 

\(\impliedby\) part. On the other way around, let us consider that $\ket{\psi}$ is an $m^{\prime}$-separable state with $m^{\prime}\ne m$. 
If $m^{\prime}\ge m+1$, there always exists at least one complete orthonormal basis locally distinguishable in at least one partition into $\le m+1$ parties, in which $\ket{\psi}$ exhibits zero coherence. 
Therefore, this satisfies (ii), but violates (i). Again, for $m^{\prime}<m$, $\ket{\psi}$ is quantum coherent with respect to all complete orthonormal bases locally distinguishable in an arbitrary partition into $m$ as well as $m+1$ parties. This obeys (i), but violates (ii). Hence, for $m^{\prime}$-separable states for $m^{\prime} \ne m$, both the conditions (i) and (ii) can not be satisfied at the same time.
\hfill \(\blacksquare\)\\


Therefore, \textit{Theorem 1} gives a necessary and sufficient condition for an $n$-party pure quantum state to be $m$-separable.
A similar theorem can be stated for mixed states.\\

\noindent \textit{Theorem 2: Any $n$-partite quantum state $\rho$, possibly mixed, is $m$-separable on the  $\mathbb{C}^{d_{1}}\otimes\mathbb{C}^{d_{2}}\otimes.....\otimes\mathbb{C}^{d_{n}}$ Hilbert space, if and only if (i) it has a nonzero quantum coherence with respect to all complete orthonormal bases that are locally distinguishable in an arbitrary partition into $m+1$ parties and (ii) exhibits a zero quantum coherence with respect to a class of complete orthonormal bases locally distinguishable in an arbitrary partition into $m$ parties.}\par
\textit{Proof:} \(\implies\) part. Let $\rho_{m\textit{-sep}}$ be an $n$-partite, $m$-separable mixed state on the Hilbert space $\mathbb{C}^{d_{1}}\otimes\mathbb{C}^{d_{2}}\otimes \ldots \otimes\mathbb{C}^{d_{n}}$. According to the definition given in Eq.~(\ref{m-sep_mixed}), there exists at least one pure-state decomposition of $\rho_{m\textit{-sep}}$, which contains at least one $m$-separable pure state, $\ket{\psi^i_{m\textit{-sep}}}$, and all other pure states in that decomposition are $m^{\prime}$-separable with $m^{\prime}\ge m$. Moreover, there is no pure state decomposition of $\rho_{m\textit{-sep}}$ containing $\tilde{m}$-separable pure states with $\tilde{m}\ge m+1$.
So, if we consider a set of all complete orthonormal bases $\{B_L^{m+1}\}$, every element of which is locally distinguishable in an arbitrary partition into $m+1$ parties, then 
$\ket{\psi^{i}_{m\textit{-sep}}}$ 
exhibits nonzero quantum coherence with respect to all elements of that set. Hence, $\rho_{m\textit{-sep}}$ is quantum coherent with respect to the class of bases $\{B_L^{m+1}\}$ according to the definition of coherence for a mixed state given in~(\ref{rel_entro_coher_mixed}).

On the other hand, suppose $\{B_L^m\}$ be the set of all complete orthonormal bases, that are locally distinguishable in an arbitrary partition of the $n$ parties into $m$ clusters. Now, let us suppose that $\ket{\psi^{i}_{m\textit{-sep}}}$ is $m$-separable in the $\mathcal{A}\equiv A_1:A_2:\ldots:A_m$ partition. Hence, if we construct an $m$-separable multi-orthonormal product basis with $\ket{\psi^{i}_{m\textit{-sep}}}$, it will be an element of the set $\{B_L^m\}$ and $\ket{\psi^{i}_{m\textit{-sep}}}$ is incoherent in that basis. Now, for another constituent pure state of $\rho_{m\textit{-sep}}$, $\ket{\psi^{i}_{m^{\prime}\textit{-sep}}}$, that is $m^{\prime}$-separable with $m^{\prime}\ge m$, 
we can again construct an $m$-separable multi-orthonormal product basis that is locally distinguishable in a partition into $m$ parties, which can be the same as or different from $\mathcal{A}$. Each of these constructed bases is an element of the set $\{B_L^m\}$ and the corresponding $\ket{\psi^{i}_{m^{\prime}\textit{-sep}}}$ are incoherent with respect to that basis. So, $\rho_{m\textit{-sep}}$ can be written as a convex sum of pure states of the same Hilbert space, with zero minimal quantum coherence when optimised over the set of bases $\{B_L^m\}$.
Thus, from the definition of quantum coherence for a mixed state (see~(\ref{rel_entro_coher_mixed})),  
$\rho_{m\textit{-sep}}$ is incoherent with respect to the set of bases $\{B_{L}^m\}$.

\(\impliedby\) part. For an $n$-partite state being $m^{\prime}$-separable, $\rho_{m^{\prime}\textit{-sep}}$, for $m^{\prime} \ne m$, both the conditions (i) and (ii) can not be satisfied simultaneously as in \textit{Theorem 1}. This completes the proof.
\hfill \(\blacksquare\)\\

We now move over to a characterisation of $k$-partite entangled states.\\


\noindent\textit{Theorem 3: An $n$-partite pure quantum state in the Hilbert space $\mathbb{C}^{d_{1}}\otimes\mathbb{C}^{d_{2}}\otimes\ldots \otimes\mathbb{C}^{d_{n}}$ is $k$-partite entangled if and only if (i) there exist at least one $k$-partite substate, that is quantum coherent with respect to all $k$-partite complete orthonormal bases, locally distinguishable in an arbitrary bipartition of the $k$ parties and (ii) each of the existing $\tilde{k}$-partite substates of the $n$-party state with $\tilde{k}\ge k+1$ is incoherent with respect to at least one $\tilde{k}$-partite complete orthonormal basis, locally distinguishable in at least one bipartition of the $\tilde{k}$ parties. 
}  

\textit{Proof:} \(\implies\) part. Let $\ket{\psi}$ be an $n$-partite pure quantum state in the Hilbert space $\mathbb{C}^{d_{1}}\otimes\mathbb{C}^{d_{2}}\otimes\mathbb{C}^{d_{3}}......\otimes\mathbb{C}^{d_{n}}$, which is $k$-partite entangled. Therefore, there exist at least one $k$-partite substate in $\ket{\psi}$, say $\ket{\psi_k} \in \mathbb{C}^{d_k^{\prime}}$, which is genuinely $k$-party entangled. 
Let us now take the set of all complete orthonormal bases $\{B_L^{k}\}$ in the Hilbert space $\mathbb{C}^{d_k^{\prime}}$, that are locally distinguishable in an arbitrary bipartition of the $k$ parties. Suppose, for one element of $\{B_L^{k}\}$, the partition be $A_1:A_{k-1}$. As $\ket{\psi_k}$ is entangled across all bipartitions of the $k$ parties, it is quantum coherent with respect to this basis. Similarly, for other elements of $\{B_L^{k}\}$ having local distinguishability in any other bipartition of the $k$ parties, $\ket{\psi_k}$ is quantum coherent in that basis also. So, $\ket{\psi_k}$ is quantum coherent with respect to all elements of the set $\{B_L^k\}$. 

Now, suppose $\{B_L^{\tilde{k}}\}$ be a set of all $\tilde{k}$-partite complete othonormal bases in the Hilbert space $\mathbb{C}^{d_{\tilde{k}}^{\prime}}$, locally distinguishable in an arbitrary bipartition into $\tilde{k}$ parties with $\tilde{k}\ge k+1$. An $n$-partite pure quantum state, which is $k$-partite entangled, does not contain a $\ge (k+1)$-partite genuinely entangled substate and hence all the $\tilde{k}$-partite substates of $\ket{\psi}$ are not genuinely entangled. It means they are separable in at least one bipartition into $\tilde{k}$ parties. So, for each of the $\tilde{k}$-partite substates, there exists at least one element in the set $\{B_L^{\tilde{k}}\}$ with respect to which the corresponding $\tilde{k}$-partite substate will be incoherent. 

\(\impliedby\) part. On the other way round, let us consider that the state $\ket{\psi}$ is $k^{\prime}$-partite entangled, where $k^{\prime}\ne k$. If $k^{\prime}\ge k+1$, there may exist a pure substate of $\ket{\psi}$ which is genuinely $k$-partite entangled and hence quantum coherent with respect to all complete orthonormal bases locally distinguishable in an arbitrary bipartition of the $k$ parties, but there also exists at least one $k^{\prime}$-partite substate which is quantum coherent with respect to all complete orthonormal bases locally distinguishable in an arbitrary bipartition of the $k^{\prime}$ parties. This satisfies condition (i), but violates condition (ii). For $k^{\prime}< k$, there is no $k$-partite pure substate of $\ket{\psi}$ which is quantum coherent with respect to all complete orthonormal bases locally distinguishable in an arbitrary bipartition of the $k$ parties as there is no $k$-partite genuinely entangled pure state. So, this violates condition (i). Therefore, for $k^{\prime}\ne k$, both the conditions (i) and (ii) can not be satisfied. This completes the proof.

\textit{Remark:} The proof of \textit{Theorem 3} is an extension of \textit{Theorem 8} of~\cite{relation_ent-coh} for a boarder spectrum of multiparty entanglement. There is a typo in the statement of \textit{Theorem 8} in~\cite{relation_ent-coh}. The correct statement is the following.\\
\\
\textit{A multiparty pure state in $\mathbb{C}^{d_{1}}\otimes\mathbb{C}^{d_{2}}\otimes\mathbb{C}^{d_{3}}......\otimes\mathbb{C}^{d_{m}}$ is genuinely multiparty entangled if and only if it is quantum coherent with respect to all complete orthonormal bases that are locally distinguishable in an arbitrary bipartition of the m parties.}\\
\\
The proof is similar with the proof of condition (i) of \textit{Theorem 3} of this paper.\hfill \(\blacksquare\)\\ 

An extension to mixed states is in the following theorem.\\
\noindent\textit{Theorem 4: An $n$-partite quantum state $\rho$,  possibly mixed, on the $\mathbb{C}^{d_{1}}\otimes\mathbb{C}^{d_{2}}\otimes.....\otimes\mathbb{C}^{d_{n}}$ Hilbert space, is $K$-partite entangled if and only if (1) there exist at least one pure state decomposition of $\rho$, in which at least one $n$-partite pure state obeys the conditions (i) and (ii) of Theorem 3 
for 
$k=K$,
and no other pure state in that decomposition obey 
either of 
the same conditions for  \(k\ge K+1\),
and (2) there is no other pure state decomposition of $\rho$, in which 
all
pure states satisfy (i) and (ii) of Theorem 3 for 
$k \le K-1$.}\\
\\

\textit{Proof:} The proof of this theorem can be obtained by applying the arguments of the proof of \textit{Theorem 3} by keeping in mind all the three conditions of a mixed state to be $k$-partite entangled, discussed in the previous section.
\hfill \(\blacksquare\)\\

The theorems discussed above characterise
$m$-separability and $k$-partite entanglement in terms of coherence and local distinguishability, and also distinguish different
types of multipartite entanglement. This approach offers a different perspective on entanglement detection, where coherence-based measures may be useful in certain contexts~\cite{Streltsov2015,Yao2015,Napoli2016,Cimini2019}.
 
For bipartite scenarios, the minimal relative entropy of quantum coherence was found to be equal with the entanglement of formation of a state in~\cite{relation_ent-coh}. Below we show that for the multiparty case, the square of minimal fidelity-based quantum coherence turns out to be equal to the geometric measure of entanglement.\\

\noindent\textit{Theorem 5: The square of the minimum fidelity-based quantum coherence
of an $n$-partite pure entangled state, 
with respect to an optimal complete orthonormal basis locally distinguishable in a partition into $m$-parties, is equal to the geometric measure of entanglement, which quantifies the $m$-inseparability of the $n$-partite entangled state. For an $n$-partite mixed state, the geometric measure of entanglement is equal to the convex sum of the square of minimal fidelity-based quantum coherence for pure states of the same Hilbert space.
}    
     
     
\textit{Proof:} Let $\ket{\psi_n^{m^{\prime}}}$ be an $n$-partite $m^{\prime}$-separable pure entangled state, 
where $m^{\prime}< m$, 
in the Hilbert space \(\mathbb{C}_G \equiv \mathbb{C}^{d_{1}}\otimes\mathbb{C}^{d_{2}}\otimes \ldots \otimes\mathbb{C}^{d_{n}}\). The \(GM_m\) 
of the state, according to the definition given in Eq.~(\ref{GM_pure}), is
\begin{equation}
GM_m(\ket{\psi_{n}^{m^{\prime}}})=\min_{\ket{P}}(1-| \braket{P|\psi_n^{m^{\prime}}}|^2),
\end{equation}
where $\ket{P}$ belongs to the set of all $m$-separable states. 
Let the optimal $\ket{P}$, that minimises the above quantity be  $\ket{P_m}=\ket{P^{\prime}_{1}}\otimes\ket{P^{\prime}_{2}}\otimes \ldots \otimes\ket{P^{\prime}_{m}}$. Let the corresponding partition into Hilbert spaces of \(\mathbb{C}_G\) be \(\mathbb{C}^{\tilde{d}_{1}}\otimes\mathbb{C}^{\tilde{d}_{2}}\otimes \ldots \otimes\mathbb{C}^{\tilde{d}_{m}}\), with \(\prod_{i=1}^n d_i = \prod_{i=1}^m \tilde{d}_i\). Therefore we have,
\begin{equation}
GM_m( \ket{\psi_n^{m^{\prime}}})=1-| \braket{P_m|\psi_{n}^{m^{\prime}}}|^2.
\end{equation} 
Now, using this $\ket{P_m}$, we can construct an $m$-separable multi-orthonormal product basis, $B_L^m$, that is LOCC distinguishable in 
\(\mathbb{C}^{\tilde{d}_{1}}\otimes\mathbb{C}^{\tilde{d}_{2}}\otimes \ldots \otimes\mathbb{C}^{\tilde{d}_{m}}\). 
Clearly, $\ket{\psi_{n}^{m^{\prime}}}$ is not an
element of $B_L^m$, as it is $m^{\prime}$-separable with $m^{\prime}<m$,
and hence will have nonzero quantum coherence with respect to that basis. Note that $\ket{P_m}$ can also be interpreted as the closest $m$-separable pure state to $\ket{\psi_{n}^{m^{\prime}}}$ that is incoherent with respect to the basis $B_L^m$. So, basically $GM_m(\ket{\psi_n^{m^{\prime}}})$ measures the minimum distance of the state $\ket{\psi_n^{m^{\prime}}}$ from the nearest pure incoherent state with respect to an optimal basis, optimised over the set of all complete orthonormal bases, $\{B_L^m\}$,  that are LOCC distinguishable in any arbitrary partition into  $m$ parties. This is equal to the square of the minimum of fidelity-based coherence measure defined in Eq.~(\ref{fid_coher}). So, $GM_m(\ket{\psi_n^{m^{\prime}}})=(C_{B_L^m}^F(\ket{\psi_n^{m^{\prime}}}))^2$,
where $C_{B_L^m}^F(\ket{\psi_n^{m^{\prime}}})$ is minimised over all the bases in the set $\{B_L^m\}$. Moreover, for an $m$-separable pure state, the $GM_m$ measure will be zero and also the minimum fidelity based coherence measure is zero as we can form a basis with that state, which will be an element of the set $\{B_L^m\}$.
So, here we are able to relate quantum coherence with the geometric measure of entanglement for an $n$-partite pure quantum state. In the case of detecting genuine multipartite entanglement of the considered state, $\ket{P}$ has to be the set of all biseparable pure states. The connection between the fidelity-based quantum coherence measure 
with the $GGM$ still holds in that case.
A convex roof approach for the extension to an $n$-partite $m^{\prime}$-separable mixed states, $\rho_n^{m^{\prime}}$, leads to the relation of the same, given by
\begin{equation}
 GM_m(\rho_n^{m^{\prime}})=\min_{\{p_i,\ket{\psi_i}\}} \sum_i p_i  (C_{B_L^m}^{F}(\ket{\psi^i}))^2.
\end{equation}
The outer minimisation is over all the pure state decompositions of $\rho_n^{m^{\prime}}=\sum_i p_i\ket{\psi^{i}}\bra{\psi^{i}}$.

\textit{Remark:} This theorem is valid only for $m^{\prime}$-separable states with $m^{\prime}\le m$. For $m^{\prime} > m$, the optimal value of quantum coherence is always zero, as all $m^{\prime}$-separable states 
are incoherent with respect to at least one complete orthonormal basis locally distinguishable in at least one partition into $m$ parties. On the other hand, according to the definition of $m$-separability taken in this paper, the geometric measure of entanglement, $GM_m$, will exhibit nonzero values for $m^{\prime}> m$, as in this case, $\ket{P_m}$ will be different from $\ket{\psi_n^{m^{\prime}}}$. \hfill \(\blacksquare\)\\

For pure states, this theorem provides an equivalence between $GM_m$ and the fidelity-based coherence measure, ensuring that the square of the coherence measure, when minimised over all locally distinguishable $m$-separable bases, precisely captures the $GM_m$. This result is significant because it establishes an explicit connection between two independently motivated resource measures—one from coherence theory and the other from entanglement theory. 
Moreover, it ensures that this connection, initially formulated for pure states, extends rigorously to mixed states even when coherence is measured in any complete orthonormal basis that is locally distinguishable under a $m$-partite partition. A connection between the geometric measure of entanglement and a fidelity-based measure of quantum coherence was already established in Ref.~\cite{Streltsov_2018,HHO2003}, highlighting the role of purity in bounding these quantum resources. Our work extends this discussion by explicitly analysing different types of partially entangled states, offering a deeper insight into the limitations and trade-offs between coherence and entanglement.
\section{CONCLUSION}
\label{sec:con}
In this paper, we established an inter-relation between the two resources, quantum entanglement and quantum coherence for multiparty scenarios. Understanding, discerning, and computing detection and quantification of entanglement for multiparty cases are of critical importance in efficient usage of quantum devices and for better comprehension of cooperative quantum phenomena. Therefore, it is fruitful to relate the multipartite entanglement of a system with other physical quantities like quantum coherence, which are \emph{a priori} unrelated to the former. 
We found that $m$-separability and $k$-partite entanglement of pure or mixed multiparty quantum states of arbitrary dimensions and arbitrary number of parties are connected qualitatively as well as quantitatively to quantum coherence of the same states in certain locally distinguishable bases. In particular, we 
found a relation between the geometric measure of $m$-inseparable entanglement and the minimum fidelity-based quantum coherence, with respect to an optimal locally distinguishable basis in a certain partition into $m$ parties. 
\section{ACKNOWLEGEMENTS}
This research was supported in part by the `INFOSYS scholarship for senior students'.  We also acknowledge partial support from the Department of Science and Technology, Government of India, through the QuEST grant with Grant No. DST/ICPS/QUST/Theme-3/2019/120. A. G. acknowledges support from the Alexander von Humboldt Foundation.
 
 \bibliography{qm} 

\begin{thebibliography}{113}%
\makeatletter
\providecommand \@ifxundefined [1]{%
 \@ifx{#1\undefined}
}%
\providecommand \@ifnum [1]{%
 \ifnum #1\expandafter \@firstoftwo
 \else \expandafter \@secondoftwo
 \fi
}%
\providecommand \@ifx [1]{%
 \ifx #1\expandafter \@firstoftwo
 \else \expandafter \@secondoftwo
 \fi
}%
\providecommand \natexlab [1]{#1}%
\providecommand \enquote  [1]{``#1''}%
\providecommand \bibnamefont  [1]{#1}%
\providecommand \bibfnamefont [1]{#1}%
\providecommand \citenamefont [1]{#1}%
\providecommand \href@noop [0]{\@secondoftwo}%
\providecommand \href [0]{\begingroup \@sanitize@url \@href}%
\providecommand \@href[1]{\@@startlink{#1}\@@href}%
\providecommand \@@href[1]{\endgroup#1\@@endlink}%
\providecommand \@sanitize@url [0]{\catcode `\\12\catcode `\$12\catcode
  `\&12\catcode `\#12\catcode `\^12\catcode `\_12\catcode `\%12\relax}%
\providecommand \@@startlink[1]{}%
\providecommand \@@endlink[0]{}%
\providecommand \url  [0]{\begingroup\@sanitize@url \@url }%
\providecommand \@url [1]{\endgroup\@href {#1}{\urlprefix }}%
\providecommand \urlprefix  [0]{URL }%
\providecommand \Eprint [0]{\href }%
\providecommand \doibase [0]{https://doi.org/}%
\providecommand \selectlanguage [0]{\@gobble}%
\providecommand \bibinfo  [0]{\@secondoftwo}%
\providecommand \bibfield  [0]{\@secondoftwo}%
\providecommand \translation [1]{[#1]}%
\providecommand \BibitemOpen [0]{}%
\providecommand \bibitemStop [0]{}%
\providecommand \bibitemNoStop [0]{.\EOS\space}%
\providecommand \EOS [0]{\spacefactor3000\relax}%
\providecommand \BibitemShut  [1]{\csname bibitem#1\endcsname}%
\let\auto@bib@innerbib\@empty
\bibitem [{\citenamefont {Horodecki}\ \emph {et~al.}(2009)\citenamefont
  {Horodecki}, \citenamefont {Horodecki}, \citenamefont {Horodecki},\ and\
  \citenamefont {Horodecki}}]{ent-review}%
  \BibitemOpen
  \bibfield  {author} {\bibinfo {author} {\bibfnamefont {R.}~\bibnamefont
  {Horodecki}}, \bibinfo {author} {\bibfnamefont {P.}~\bibnamefont
  {Horodecki}}, \bibinfo {author} {\bibfnamefont {M.}~\bibnamefont
  {Horodecki}},\ and\ \bibinfo {author} {\bibfnamefont {K.}~\bibnamefont
  {Horodecki}},\ }\bibfield  {title} {\bibinfo {title} {Quantum entanglement},\
  }\href {https://link.aps.org/doi/10.1103/RevModPhys.81.865} {\bibfield
  {journal} {\bibinfo  {journal} {Rev. Mod. Phys.}\ }\textbf {\bibinfo {volume}
  {81}},\ \bibinfo {pages} {865} (\bibinfo {year} {2009})}\BibitemShut
  {NoStop}%
\bibitem [{\citenamefont {G\"{u}hne}\ and\ \citenamefont
  {T\'{o}th}(2009)}]{GUHNE20091}%
  \BibitemOpen
  \bibfield  {author} {\bibinfo {author} {\bibfnamefont {O.}~\bibnamefont
  {G\"{u}hne}}\ and\ \bibinfo {author} {\bibfnamefont {G.}~\bibnamefont
  {T\'{o}th}},\ }\bibfield  {title} {\bibinfo {title} {Entanglement
  detection},\ }\href
  {https://doi.org/https://doi.org/10.1016/j.physrep.2009.02.004} {\bibfield
  {journal} {\bibinfo  {journal} {Phys. Rep.}\ }\textbf {\bibinfo {volume}
  {474}},\ \bibinfo {pages} {1} (\bibinfo {year} {2009})}\BibitemShut {NoStop}%
\bibitem [{\citenamefont {Das}\ \emph {et~al.}(2019)\citenamefont {Das},
  \citenamefont {Chanda}, \citenamefont {Lewenstein}, \citenamefont {Sanpera},
  \citenamefont {Sen(De)},\ and\ \citenamefont {Sen}}]{Das2019}%
  \BibitemOpen
  \bibfield  {author} {\bibinfo {author} {\bibfnamefont {S.}~\bibnamefont
  {Das}}, \bibinfo {author} {\bibfnamefont {T.}~\bibnamefont {Chanda}},
  \bibinfo {author} {\bibfnamefont {M.}~\bibnamefont {Lewenstein}}, \bibinfo
  {author} {\bibfnamefont {A.}~\bibnamefont {Sanpera}}, \bibinfo {author}
  {\bibfnamefont {A.}~\bibnamefont {Sen(De)}},\ and\ \bibinfo {author}
  {\bibfnamefont {U.}~\bibnamefont {Sen}},\ }\href@noop {} {\emph {\bibinfo
  {title} {The separability versus entanglement problem, in Quantum
  Information: From Foun- dations to Quantum Technology Applications, 2nd ed.,
  edited by D. Bruß and G. Leuchs}}}\ (\bibinfo  {publisher} {Wiley,
  Weinheim},\ \bibinfo {year} {2019})\BibitemShut {NoStop}%
\bibitem [{\citenamefont {\r{A}berg}()}]{Aberg2006}%
  \BibitemOpen
  \bibfield  {author} {\bibinfo {author} {\bibfnamefont {J.}~\bibnamefont
  {\r{A}berg}},\ }\bibfield  {title} {\bibinfo {title} {Quantifying
  superposition},\ }\href@noop {} {\bibinfo  {journal}
  {arXiv:quant-ph/0612146}\ }\BibitemShut {NoStop}%
\bibitem [{\citenamefont {Baumgratz}\ \emph {et~al.}(2014)\citenamefont
  {Baumgratz}, \citenamefont {Cramer},\ and\ \citenamefont {Plenio}}]{coh}%
  \BibitemOpen
\bibfield  {journal} {  }\bibfield  {author} {\bibinfo {author} {\bibfnamefont
  {T.}~\bibnamefont {Baumgratz}}, \bibinfo {author} {\bibfnamefont
  {M.}~\bibnamefont {Cramer}},\ and\ \bibinfo {author} {\bibfnamefont {M.~B.}\
  \bibnamefont {Plenio}},\ }\bibfield  {title} {\bibinfo {title} {Quantifying
  coherence},\ }\href {https://link.aps.org/doi/10.1103/PhysRevLett.113.140401}
  {\bibfield  {journal} {\bibinfo  {journal} {Phys. Rev. Lett.}\ }\textbf
  {\bibinfo {volume} {113}},\ \bibinfo {pages} {140401} (\bibinfo {year}
  {2014})}\BibitemShut {NoStop}%
\bibitem [{\citenamefont {Winter}\ and\ \citenamefont
  {Yang}(2016)}]{Winter2016}%
  \BibitemOpen
  \bibfield  {author} {\bibinfo {author} {\bibfnamefont {A.}~\bibnamefont
  {Winter}}\ and\ \bibinfo {author} {\bibfnamefont {D.}~\bibnamefont {Yang}},\
  }\bibfield  {title} {\bibinfo {title} {Operational resource theory of
  coherence},\ }\href {https://doi.org/10.1103/PhysRevLett.116.120404}
  {\bibfield  {journal} {\bibinfo  {journal} {Phys. Rev. Lett.}\ }\textbf
  {\bibinfo {volume} {116}},\ \bibinfo {pages} {120404} (\bibinfo {year}
  {2016})}\BibitemShut {NoStop}%
\bibitem [{\citenamefont {Streltsov}\ \emph {et~al.}(2017)\citenamefont
  {Streltsov}, \citenamefont {Adesso},\ and\ \citenamefont
  {Plenio}}]{Streltsov2017}%
  \BibitemOpen
  \bibfield  {author} {\bibinfo {author} {\bibfnamefont {A.}~\bibnamefont
  {Streltsov}}, \bibinfo {author} {\bibfnamefont {G.}~\bibnamefont {Adesso}},\
  and\ \bibinfo {author} {\bibfnamefont {M.~B.}\ \bibnamefont {Plenio}},\
  }\bibfield  {title} {\bibinfo {title} {Colloquium: Quantum coherence as a
  resource},\ }\href {https://doi.org/10.1103/RevModPhys.89.041003} {\bibfield
  {journal} {\bibinfo  {journal} {Rev. Mod. Phys.}\ }\textbf {\bibinfo {volume}
  {89}},\ \bibinfo {pages} {041003} (\bibinfo {year} {2017})}\BibitemShut
  {NoStop}%
\bibitem [{\citenamefont {Bennett}\ and\ \citenamefont
  {Wiesner}(1992)}]{supcod}%
  \BibitemOpen
  \bibfield  {author} {\bibinfo {author} {\bibfnamefont {C.~H.}\ \bibnamefont
  {Bennett}}\ and\ \bibinfo {author} {\bibfnamefont {S.~J.}\ \bibnamefont
  {Wiesner}},\ }\bibfield  {title} {\bibinfo {title} {Communication via one-
  and two-particle operators on einstein-podolsky-rosen states},\ }\href
  {https://link.aps.org/doi/10.1103/PhysRevLett.69.2881} {\bibfield  {journal}
  {\bibinfo  {journal} {Phys. Rev. Lett.}\ }\textbf {\bibinfo {volume} {69}},\
  \bibinfo {pages} {2881} (\bibinfo {year} {1992})}\BibitemShut {NoStop}%
\bibitem [{\citenamefont {Guo}\ \emph {et~al.}(2019)\citenamefont {Guo},
  \citenamefont {Liu}, \citenamefont {Li},\ and\ \citenamefont
  {Guo}}]{Guo2019}%
  \BibitemOpen
  \bibfield  {author} {\bibinfo {author} {\bibfnamefont {Y.}~\bibnamefont
  {Guo}}, \bibinfo {author} {\bibfnamefont {B.-H.}\ \bibnamefont {Liu}},
  \bibinfo {author} {\bibfnamefont {C.-F.}\ \bibnamefont {Li}},\ and\ \bibinfo
  {author} {\bibfnamefont {G.-C.}\ \bibnamefont {Guo}},\ }\bibfield  {title}
  {\bibinfo {title} {Advances in quantum dense coding},\ }\href
  {https://doi.org/https://doi.org/10.1002/qute.201900011} {\bibfield
  {journal} {\bibinfo  {journal} {Advanced Quantum Technologies}\ }\textbf
  {\bibinfo {volume} {2}},\ \bibinfo {pages} {1900011} (\bibinfo {year}
  {2019})}\BibitemShut {NoStop}%
\bibitem [{\citenamefont {Bennett}\ \emph {et~al.}(1993)\citenamefont
  {Bennett}, \citenamefont {Brassard}, \citenamefont {Cr\'epeau}, \citenamefont
  {Jozsa}, \citenamefont {Peres},\ and\ \citenamefont {Wootters}}]{tele}%
  \BibitemOpen
  \bibfield  {author} {\bibinfo {author} {\bibfnamefont {C.~H.}\ \bibnamefont
  {Bennett}}, \bibinfo {author} {\bibfnamefont {G.}~\bibnamefont {Brassard}},
  \bibinfo {author} {\bibfnamefont {C.}~\bibnamefont {Cr\'epeau}}, \bibinfo
  {author} {\bibfnamefont {R.}~\bibnamefont {Jozsa}}, \bibinfo {author}
  {\bibfnamefont {A.}~\bibnamefont {Peres}},\ and\ \bibinfo {author}
  {\bibfnamefont {W.~K.}\ \bibnamefont {Wootters}},\ }\bibfield  {title}
  {\bibinfo {title} {Teleporting an unknown quantum state via dual classical
  and einstein-podolsky-rosen channels},\ }\href
  {https://link.aps.org/doi/10.1103/PhysRevLett.70.1895} {\bibfield  {journal}
  {\bibinfo  {journal} {Phys. Rev. Lett.}\ }\textbf {\bibinfo {volume} {70}},\
  \bibinfo {pages} {1895} (\bibinfo {year} {1993})}\BibitemShut {NoStop}%
\bibitem [{\citenamefont {Pirandola}\ \emph {et~al.}(2015)\citenamefont
  {Pirandola}, \citenamefont {Eisert}, \citenamefont {Weedbrook}, \citenamefont
  {Furusawa},\ and\ \citenamefont {Braunstein}}]{Pirandola2015}%
  \BibitemOpen
  \bibfield  {author} {\bibinfo {author} {\bibfnamefont {S.}~\bibnamefont
  {Pirandola}}, \bibinfo {author} {\bibfnamefont {J.}~\bibnamefont {Eisert}},
  \bibinfo {author} {\bibfnamefont {C.}~\bibnamefont {Weedbrook}}, \bibinfo
  {author} {\bibfnamefont {A.}~\bibnamefont {Furusawa}},\ and\ \bibinfo
  {author} {\bibfnamefont {S.~L.}\ \bibnamefont {Braunstein}},\ }\bibfield
  {title} {\bibinfo {title} {Advances in quantum teleportation},\ }\href
  {https://doi.org/10.1038/nphoton.2015.154} {\bibfield  {journal} {\bibinfo
  {journal} {Nature Photonics}\ }\textbf {\bibinfo {volume} {9}},\ \bibinfo
  {pages} {641} (\bibinfo {year} {2015})}\BibitemShut {NoStop}%
\bibitem [{\citenamefont {Liu}(2020)}]{Liu_2020}%
  \BibitemOpen
  \bibfield  {author} {\bibinfo {author} {\bibfnamefont {T.}~\bibnamefont
  {Liu}},\ }\bibfield  {title} {\bibinfo {title} {The applications and
  challenges of quantum teleportation},\ }\href
  {https://doi.org/10.1088/1742-6596/1634/1/012089} {\bibfield  {journal}
  {\bibinfo  {journal} {Journal of Physics: Conference Series}\ }\textbf
  {\bibinfo {volume} {1634}},\ \bibinfo {pages} {012089} (\bibinfo {year}
  {2020})}\BibitemShut {NoStop}%
\bibitem [{\citenamefont {Ekert}(1991)}]{qcryp}%
  \BibitemOpen
  \bibfield  {author} {\bibinfo {author} {\bibfnamefont {A.~K.}\ \bibnamefont
  {Ekert}},\ }\bibfield  {title} {\bibinfo {title} {Quantum cryptography based
  on bell's theorem},\ }\href {https://doi.org/10.1103/PhysRevLett.67.661}
  {\bibfield  {journal} {\bibinfo  {journal} {Phys. Rev. Lett.}\ }\textbf
  {\bibinfo {volume} {67}},\ \bibinfo {pages} {661} (\bibinfo {year}
  {1991})}\BibitemShut {NoStop}%
\bibitem [{\citenamefont {Gisin}\ \emph {et~al.}(2002)\citenamefont {Gisin},
  \citenamefont {Ribordy}, \citenamefont {Tittel},\ and\ \citenamefont
  {Zbinden}}]{Gisin2002}%
  \BibitemOpen
  \bibfield  {author} {\bibinfo {author} {\bibfnamefont {N.}~\bibnamefont
  {Gisin}}, \bibinfo {author} {\bibfnamefont {G.}~\bibnamefont {Ribordy}},
  \bibinfo {author} {\bibfnamefont {W.}~\bibnamefont {Tittel}},\ and\ \bibinfo
  {author} {\bibfnamefont {H.}~\bibnamefont {Zbinden}},\ }\bibfield  {title}
  {\bibinfo {title} {Quantum cryptography},\ }\href
  {https://doi.org/10.1103/RevModPhys.74.145} {\bibfield  {journal} {\bibinfo
  {journal} {Rev. Mod. Phys.}\ }\textbf {\bibinfo {volume} {74}},\ \bibinfo
  {pages} {145} (\bibinfo {year} {2002})}\BibitemShut {NoStop}%
\bibitem [{\citenamefont {Pirandola}\ \emph {et~al.}(2020)\citenamefont
  {Pirandola}, \citenamefont {Andersen}, \citenamefont {Banchi}, \citenamefont
  {Berta}, \citenamefont {Bunandar}, \citenamefont {Colbeck}, \citenamefont
  {Englund}, \citenamefont {Gehring}, \citenamefont {Lupo}, \citenamefont
  {Ottaviani}, \citenamefont {Pereira}, \citenamefont {Razavi}, \citenamefont
  {Shaari}, \citenamefont {Tomamichel}, \citenamefont {Usenko}, \citenamefont
  {Vallone}, \citenamefont {Villoresi},\ and\ \citenamefont
  {Wallden}}]{Pirandola20}%
  \BibitemOpen
  \bibfield  {author} {\bibinfo {author} {\bibfnamefont {S.}~\bibnamefont
  {Pirandola}}, \bibinfo {author} {\bibfnamefont {U.~L.}\ \bibnamefont
  {Andersen}}, \bibinfo {author} {\bibfnamefont {L.}~\bibnamefont {Banchi}},
  \bibinfo {author} {\bibfnamefont {M.}~\bibnamefont {Berta}}, \bibinfo
  {author} {\bibfnamefont {D.}~\bibnamefont {Bunandar}}, \bibinfo {author}
  {\bibfnamefont {R.}~\bibnamefont {Colbeck}}, \bibinfo {author} {\bibfnamefont
  {D.}~\bibnamefont {Englund}}, \bibinfo {author} {\bibfnamefont
  {T.}~\bibnamefont {Gehring}}, \bibinfo {author} {\bibfnamefont
  {C.}~\bibnamefont {Lupo}}, \bibinfo {author} {\bibfnamefont {C.}~\bibnamefont
  {Ottaviani}}, \bibinfo {author} {\bibfnamefont {J.~L.}\ \bibnamefont
  {Pereira}}, \bibinfo {author} {\bibfnamefont {M.}~\bibnamefont {Razavi}},
  \bibinfo {author} {\bibfnamefont {J.~S.}\ \bibnamefont {Shaari}}, \bibinfo
  {author} {\bibfnamefont {M.}~\bibnamefont {Tomamichel}}, \bibinfo {author}
  {\bibfnamefont {V.~C.}\ \bibnamefont {Usenko}}, \bibinfo {author}
  {\bibfnamefont {G.}~\bibnamefont {Vallone}}, \bibinfo {author} {\bibfnamefont
  {P.}~\bibnamefont {Villoresi}},\ and\ \bibinfo {author} {\bibfnamefont
  {P.}~\bibnamefont {Wallden}},\ }\bibfield  {title} {\bibinfo {title}
  {Advances in quantum cryptography},\ }\href
  {https://doi.org/10.1364/AOP.361502} {\bibfield  {journal} {\bibinfo
  {journal} {Adv. Opt. Photon.}\ }\textbf {\bibinfo {volume} {12}},\ \bibinfo
  {pages} {1012} (\bibinfo {year} {2020})}\BibitemShut {NoStop}%
\bibitem [{\citenamefont {Portmann}\ and\ \citenamefont
  {Renner}(2022)}]{Portmann2021}%
  \BibitemOpen
  \bibfield  {author} {\bibinfo {author} {\bibfnamefont {C.}~\bibnamefont
  {Portmann}}\ and\ \bibinfo {author} {\bibfnamefont {R.}~\bibnamefont
  {Renner}},\ }\bibfield  {title} {\bibinfo {title} {Security in quantum
  cryptography},\ }\href {https://doi.org/10.1103/RevModPhys.94.025008}
  {\bibfield  {journal} {\bibinfo  {journal} {Rev. Mod. Phys.}\ }\textbf
  {\bibinfo {volume} {94}},\ \bibinfo {pages} {025008} (\bibinfo {year}
  {2022})}\BibitemShut {NoStop}%
\bibitem [{\citenamefont {{Lukin}}(1998)}]{Lukin1998}%
  \BibitemOpen
  \bibfield  {author} {\bibinfo {author} {\bibfnamefont {M.~D.}\ \bibnamefont
  {{Lukin}}},\ }\emph {\bibinfo {title} {{Quantum coherence and interference in
  optics and laser spectroscopy}}},\ \href@noop {} {Ph.D. thesis},\ \bibinfo
  {school} {Texas A\&M University System} (\bibinfo {year} {1998})\BibitemShut
  {NoStop}%
\bibitem [{\citenamefont {Pires}\ \emph {et~al.}(2018)\citenamefont {Pires},
  \citenamefont {Silva}, \citenamefont {deAzevedo}, \citenamefont
  {Soares-Pinto},\ and\ \citenamefont {Filgueiras}}]{Pires2018}%
  \BibitemOpen
  \bibfield  {author} {\bibinfo {author} {\bibfnamefont {D.~P.}\ \bibnamefont
  {Pires}}, \bibinfo {author} {\bibfnamefont {I.~A.}\ \bibnamefont {Silva}},
  \bibinfo {author} {\bibfnamefont {E.~R.}\ \bibnamefont {deAzevedo}}, \bibinfo
  {author} {\bibfnamefont {D.~O.}\ \bibnamefont {Soares-Pinto}},\ and\ \bibinfo
  {author} {\bibfnamefont {J.~G.}\ \bibnamefont {Filgueiras}},\ }\bibfield
  {title} {\bibinfo {title} {Coherence orders, decoherence, and quantum
  metrology},\ }\href {https://doi.org/10.1103/PhysRevA.98.032101} {\bibfield
  {journal} {\bibinfo  {journal} {Phys. Rev. A}\ }\textbf {\bibinfo {volume}
  {98}},\ \bibinfo {pages} {032101} (\bibinfo {year} {2018})}\BibitemShut
  {NoStop}%
\bibitem [{\citenamefont {Castellini}\ \emph {et~al.}(2019)\citenamefont
  {Castellini}, \citenamefont {Lo~Franco}, \citenamefont {Lami}, \citenamefont
  {Winter}, \citenamefont {Adesso},\ and\ \citenamefont
  {Compagno}}]{Castellini2019}%
  \BibitemOpen
  \bibfield  {author} {\bibinfo {author} {\bibfnamefont {A.}~\bibnamefont
  {Castellini}}, \bibinfo {author} {\bibfnamefont {R.}~\bibnamefont
  {Lo~Franco}}, \bibinfo {author} {\bibfnamefont {L.}~\bibnamefont {Lami}},
  \bibinfo {author} {\bibfnamefont {A.}~\bibnamefont {Winter}}, \bibinfo
  {author} {\bibfnamefont {G.}~\bibnamefont {Adesso}},\ and\ \bibinfo {author}
  {\bibfnamefont {G.}~\bibnamefont {Compagno}},\ }\bibfield  {title} {\bibinfo
  {title} {Indistinguishability-enabled coherence for quantum metrology},\
  }\href {https://doi.org/10.1103/PhysRevA.100.012308} {\bibfield  {journal}
  {\bibinfo  {journal} {Phys. Rev. A}\ }\textbf {\bibinfo {volume} {100}},\
  \bibinfo {pages} {012308} (\bibinfo {year} {2019})}\BibitemShut {NoStop}%
\bibitem [{\citenamefont {Zhang}\ \emph {et~al.}(2019)\citenamefont {Zhang},
  \citenamefont {Bromley}, \citenamefont {Huang}, \citenamefont {Cao},
  \citenamefont {Lv}, \citenamefont {Liu}, \citenamefont {Li}, \citenamefont
  {Guo}, \citenamefont {Cianciaruso},\ and\ \citenamefont
  {Adesso}}]{Zhang2019}%
  \BibitemOpen
  \bibfield  {author} {\bibinfo {author} {\bibfnamefont {C.}~\bibnamefont
  {Zhang}}, \bibinfo {author} {\bibfnamefont {T.~R.}\ \bibnamefont {Bromley}},
  \bibinfo {author} {\bibfnamefont {Y.-F.}\ \bibnamefont {Huang}}, \bibinfo
  {author} {\bibfnamefont {H.}~\bibnamefont {Cao}}, \bibinfo {author}
  {\bibfnamefont {W.-M.}\ \bibnamefont {Lv}}, \bibinfo {author} {\bibfnamefont
  {B.-H.}\ \bibnamefont {Liu}}, \bibinfo {author} {\bibfnamefont {C.-F.}\
  \bibnamefont {Li}}, \bibinfo {author} {\bibfnamefont {G.-C.}\ \bibnamefont
  {Guo}}, \bibinfo {author} {\bibfnamefont {M.}~\bibnamefont {Cianciaruso}},\
  and\ \bibinfo {author} {\bibfnamefont {G.}~\bibnamefont {Adesso}},\
  }\bibfield  {title} {\bibinfo {title} {Demonstrating quantum coherence and
  metrology that is resilient to transversal noise},\ }\href
  {https://doi.org/10.1103/PhysRevLett.123.180504} {\bibfield  {journal}
  {\bibinfo  {journal} {Phys. Rev. Lett.}\ }\textbf {\bibinfo {volume} {123}},\
  \bibinfo {pages} {180504} (\bibinfo {year} {2019})}\BibitemShut {NoStop}%
\bibitem [{\citenamefont {Sun}\ \emph {et~al.}(2022)\citenamefont {Sun},
  \citenamefont {Liu}, \citenamefont {Wang}, \citenamefont {Hao}, \citenamefont
  {Xu}, \citenamefont {Xu}, \citenamefont {Li}, \citenamefont {Guo},
  \citenamefont {Castellini}, \citenamefont {Lami}, \citenamefont {Winter},
  \citenamefont {Adesso}, \citenamefont {Compagno},\ and\ \citenamefont
  {Franco}}]{Sun2022}%
  \BibitemOpen
  \bibfield  {author} {\bibinfo {author} {\bibfnamefont {K.}~\bibnamefont
  {Sun}}, \bibinfo {author} {\bibfnamefont {Z.-H.}\ \bibnamefont {Liu}},
  \bibinfo {author} {\bibfnamefont {Y.}~\bibnamefont {Wang}}, \bibinfo {author}
  {\bibfnamefont {Z.-Y.}\ \bibnamefont {Hao}}, \bibinfo {author} {\bibfnamefont
  {X.-Y.}\ \bibnamefont {Xu}}, \bibinfo {author} {\bibfnamefont {J.-S.}\
  \bibnamefont {Xu}}, \bibinfo {author} {\bibfnamefont {C.-F.}\ \bibnamefont
  {Li}}, \bibinfo {author} {\bibfnamefont {G.-C.}\ \bibnamefont {Guo}},
  \bibinfo {author} {\bibfnamefont {A.}~\bibnamefont {Castellini}}, \bibinfo
  {author} {\bibfnamefont {L.}~\bibnamefont {Lami}}, \bibinfo {author}
  {\bibfnamefont {A.}~\bibnamefont {Winter}}, \bibinfo {author} {\bibfnamefont
  {G.}~\bibnamefont {Adesso}}, \bibinfo {author} {\bibfnamefont
  {G.}~\bibnamefont {Compagno}},\ and\ \bibinfo {author} {\bibfnamefont
  {R.~L.}\ \bibnamefont {Franco}},\ }\bibfield  {title} {\bibinfo {title}
  {Activation of indistinguishability-based quantum coherence for enhanced
  metrological applications with particle statistics imprint},\ }\href
  {https://doi.org/10.1073/pnas.2119765119} {\bibfield  {journal} {\bibinfo
  {journal} {Proceedings of the National Academy of Sciences}\ }\textbf
  {\bibinfo {volume} {119}},\ \bibinfo {pages} {e2119765119} (\bibinfo {year}
  {2022})}\BibitemShut {NoStop}%
\bibitem [{\citenamefont {Hillery}(2016)}]{Hillery2016}%
  \BibitemOpen
  \bibfield  {author} {\bibinfo {author} {\bibfnamefont {M.}~\bibnamefont
  {Hillery}},\ }\bibfield  {title} {\bibinfo {title} {Coherence as a resource
  in decision problems: The deutsch-jozsa algorithm and a variation},\ }\href
  {https://link.aps.org/doi/10.1103/PhysRevA.93.012111} {\bibfield  {journal}
  {\bibinfo  {journal} {Phys. Rev. A}\ }\textbf {\bibinfo {volume} {93}},\
  \bibinfo {pages} {012111} (\bibinfo {year} {2016})}\BibitemShut {NoStop}%
\bibitem [{\citenamefont {Anand}\ and\ \citenamefont {Pati}()}]{Anand2016}%
  \BibitemOpen
  \bibfield  {author} {\bibinfo {author} {\bibfnamefont {N.}~\bibnamefont
  {Anand}}\ and\ \bibinfo {author} {\bibfnamefont {A.~K.}\ \bibnamefont
  {Pati}},\ }\bibfield  {title} {\bibinfo {title} {Coherence and entanglement
  monogamy in the discrete analogue of analog {G}rover search},\ }\href@noop {}
  {\bibinfo  {journal} {arXiv:1611.04542}\ }\BibitemShut {NoStop}%
\bibitem [{\citenamefont {Shi}\ \emph {et~al.}(2017)\citenamefont {Shi},
  \citenamefont {Liu}, \citenamefont {Wang}, \citenamefont {Yang},
  \citenamefont {Yang},\ and\ \citenamefont {Fan}}]{Shi2017}%
  \BibitemOpen
\bibfield  {journal} {  }\bibfield  {author} {\bibinfo {author} {\bibfnamefont
  {H.-L.}\ \bibnamefont {Shi}}, \bibinfo {author} {\bibfnamefont {S.-Y.}\
  \bibnamefont {Liu}}, \bibinfo {author} {\bibfnamefont {X.-H.}\ \bibnamefont
  {Wang}}, \bibinfo {author} {\bibfnamefont {W.-L.}\ \bibnamefont {Yang}},
  \bibinfo {author} {\bibfnamefont {Z.-Y.}\ \bibnamefont {Yang}},\ and\
  \bibinfo {author} {\bibfnamefont {H.}~\bibnamefont {Fan}},\ }\bibfield
  {title} {\bibinfo {title} {Coherence depletion in the grover quantum search
  algorithm},\ }\href {https://doi.org/10.1103/PhysRevA.95.032307} {\bibfield
  {journal} {\bibinfo  {journal} {Phys. Rev. A}\ }\textbf {\bibinfo {volume}
  {95}},\ \bibinfo {pages} {032307} (\bibinfo {year} {2017})}\BibitemShut
  {NoStop}%
\bibitem [{\citenamefont {Liu}\ \emph {et~al.}(2019)\citenamefont {Liu},
  \citenamefont {Shang},\ and\ \citenamefont {Zhang}}]{Liu2019}%
  \BibitemOpen
  \bibfield  {author} {\bibinfo {author} {\bibfnamefont {Y.-C.}\ \bibnamefont
  {Liu}}, \bibinfo {author} {\bibfnamefont {J.}~\bibnamefont {Shang}},\ and\
  \bibinfo {author} {\bibfnamefont {X.}~\bibnamefont {Zhang}},\ }\bibfield
  {title} {\bibinfo {title} {Coherence depletion in quantum algorithms},\
  }\href {https://doi.org/10.3390/e21030260} {\bibfield  {journal} {\bibinfo
  {journal} {Entropy}\ }\textbf {\bibinfo {volume} {21}},\ \bibinfo {pages}
  {260} (\bibinfo {year} {2019})}\BibitemShut {NoStop}%
\bibitem [{\citenamefont {S}\ and\ \citenamefont {Sen}()}]{Shubhalakshmi2020}%
  \BibitemOpen
  \bibfield  {author} {\bibinfo {author} {\bibfnamefont {S.}~\bibnamefont {S}}\
  and\ \bibinfo {author} {\bibfnamefont {U.}~\bibnamefont {Sen}},\ }\bibfield
  {title} {\bibinfo {title} {Noncommutative coherence and quantum phase
  estimation algorithm},\ }\href@noop {} {\bibinfo  {journal}
  {arXiv:2004.01419}\ }\BibitemShut {NoStop}%
\bibitem [{\citenamefont {Xiong}\ and\ \citenamefont {Wu}(2018)}]{Xiong_2018}%
  \BibitemOpen
\bibfield  {journal} {  }\bibfield  {author} {\bibinfo {author} {\bibfnamefont
  {C.}~\bibnamefont {Xiong}}\ and\ \bibinfo {author} {\bibfnamefont
  {J.}~\bibnamefont {Wu}},\ }\bibfield  {title} {\bibinfo {title} {Geometric
  coherence and quantum state discrimination},\ }\href
  {https://doi.org/10.1088/1751-8121/aac979} {\bibfield  {journal} {\bibinfo
  {journal} {Journal of Physics A: Mathematical and Theoretical}\ }\textbf
  {\bibinfo {volume} {51}},\ \bibinfo {pages} {414005} (\bibinfo {year}
  {2018})}\BibitemShut {NoStop}%
\bibitem [{\citenamefont {Kim}\ \emph {et~al.}(2021)\citenamefont {Kim},
  \citenamefont {Li}, \citenamefont {Kumar}, \citenamefont {Xiong},
  \citenamefont {Das}, \citenamefont {Sen}, \citenamefont {Pati},\ and\
  \citenamefont {Wu}}]{Kim2018}%
  \BibitemOpen
  \bibfield  {author} {\bibinfo {author} {\bibfnamefont {S.}~\bibnamefont
  {Kim}}, \bibinfo {author} {\bibfnamefont {L.}~\bibnamefont {Li}}, \bibinfo
  {author} {\bibfnamefont {A.}~\bibnamefont {Kumar}}, \bibinfo {author}
  {\bibfnamefont {C.}~\bibnamefont {Xiong}}, \bibinfo {author} {\bibfnamefont
  {S.}~\bibnamefont {Das}}, \bibinfo {author} {\bibfnamefont {U.}~\bibnamefont
  {Sen}}, \bibinfo {author} {\bibfnamefont {A.~K.}\ \bibnamefont {Pati}},\ and\
  \bibinfo {author} {\bibfnamefont {J.}~\bibnamefont {Wu}},\ }\bibfield
  {title} {\bibinfo {title} {Protocol for unambiguous quantum state
  discrimination using quantum coherence},\ }\href
  {https://doi.org/https://doi.org/10.26421/QIC21.11-12-2} {\bibfield
  {journal} {\bibinfo  {journal} {Quantum Information and Computation}\
  }\textbf {\bibinfo {volume} {21}},\ \bibinfo {pages} {0931} (\bibinfo {year}
  {2021})}\BibitemShut {NoStop}%
\bibitem [{\citenamefont {Lostaglio}\ \emph {et~al.}(2015)\citenamefont
  {Lostaglio}, \citenamefont {Korzekwa}, \citenamefont {Jennings},\ and\
  \citenamefont {Rudolph}}]{Lostaglio2015}%
  \BibitemOpen
  \bibfield  {author} {\bibinfo {author} {\bibfnamefont {M.}~\bibnamefont
  {Lostaglio}}, \bibinfo {author} {\bibfnamefont {K.}~\bibnamefont {Korzekwa}},
  \bibinfo {author} {\bibfnamefont {D.}~\bibnamefont {Jennings}},\ and\
  \bibinfo {author} {\bibfnamefont {T.}~\bibnamefont {Rudolph}},\ }\bibfield
  {title} {\bibinfo {title} {Quantum coherence, time-translation symmetry, and
  thermodynamics},\ }\href {https://doi.org/10.1103/PhysRevX.5.021001}
  {\bibfield  {journal} {\bibinfo  {journal} {Phys. Rev. X}\ }\textbf {\bibinfo
  {volume} {5}},\ \bibinfo {pages} {021001} (\bibinfo {year}
  {2015})}\BibitemShut {NoStop}%
\bibitem [{\citenamefont {Goold}\ \emph {et~al.}(2016)\citenamefont {Goold},
  \citenamefont {Huber}, \citenamefont {Riera}, \citenamefont {del Rio},\ and\
  \citenamefont {Skrzypczyk}}]{thermo}%
  \BibitemOpen
  \bibfield  {author} {\bibinfo {author} {\bibfnamefont {J.}~\bibnamefont
  {Goold}}, \bibinfo {author} {\bibfnamefont {M.}~\bibnamefont {Huber}},
  \bibinfo {author} {\bibfnamefont {A.}~\bibnamefont {Riera}}, \bibinfo
  {author} {\bibfnamefont {L.}~\bibnamefont {del Rio}},\ and\ \bibinfo {author}
  {\bibfnamefont {P.}~\bibnamefont {Skrzypczyk}},\ }\bibfield  {title}
  {\bibinfo {title} {The role of quantum information in
  thermodynamics{\textemdash}{A} topical review},\ }\href
  {https://doi.org/10.1088/1751-8113/49/14/143001} {\bibfield  {journal}
  {\bibinfo  {journal} {Journal of Physics A: Mathematical and Theoretical}\
  }\textbf {\bibinfo {volume} {49}},\ \bibinfo {pages} {143001} (\bibinfo
  {year} {2016})}\BibitemShut {NoStop}%
\bibitem [{\citenamefont {Misra}\ \emph {et~al.}(2016)\citenamefont {Misra},
  \citenamefont {Singh}, \citenamefont {Bhattacharya},\ and\ \citenamefont
  {Pati}}]{Misra2016}%
  \BibitemOpen
  \bibfield  {author} {\bibinfo {author} {\bibfnamefont {A.}~\bibnamefont
  {Misra}}, \bibinfo {author} {\bibfnamefont {U.}~\bibnamefont {Singh}},
  \bibinfo {author} {\bibfnamefont {S.}~\bibnamefont {Bhattacharya}},\ and\
  \bibinfo {author} {\bibfnamefont {A.~K.}\ \bibnamefont {Pati}},\ }\bibfield
  {title} {\bibinfo {title} {Energy cost of creating quantum coherence},\
  }\href {https://doi.org/10.1103/PhysRevA.93.052335} {\bibfield  {journal}
  {\bibinfo  {journal} {Phys. Rev. A}\ }\textbf {\bibinfo {volume} {93}},\
  \bibinfo {pages} {052335} (\bibinfo {year} {2016})}\BibitemShut {NoStop}%
\bibitem [{\citenamefont {Lostaglio}\ \emph {et~al.}(2017)\citenamefont
  {Lostaglio}, \citenamefont {Jennings},\ and\ \citenamefont
  {Rudolph}}]{Lostaglio_2017}%
  \BibitemOpen
  \bibfield  {author} {\bibinfo {author} {\bibfnamefont {M.}~\bibnamefont
  {Lostaglio}}, \bibinfo {author} {\bibfnamefont {D.}~\bibnamefont
  {Jennings}},\ and\ \bibinfo {author} {\bibfnamefont {T.}~\bibnamefont
  {Rudolph}},\ }\bibfield  {title} {\bibinfo {title} {Thermodynamic resource
  theories, non-commutativity and maximum entropy principles},\ }\href
  {https://doi.org/10.1088/1367-2630/aa617f} {\bibfield  {journal} {\bibinfo
  {journal} {New Journal of Physics}\ }\textbf {\bibinfo {volume} {19}},\
  \bibinfo {pages} {043008} (\bibinfo {year} {2017})}\BibitemShut {NoStop}%
\bibitem [{\citenamefont {Huelga}\ and\ \citenamefont
  {Plenio}(2013)}]{Huelga2013}%
  \BibitemOpen
  \bibfield  {author} {\bibinfo {author} {\bibfnamefont {S.}~\bibnamefont
  {Huelga}}\ and\ \bibinfo {author} {\bibfnamefont {M.}~\bibnamefont
  {Plenio}},\ }\bibfield  {title} {\bibinfo {title} {Vibrations, quanta and
  biology},\ }\href {https://doi.org/10.1080/00405000.2013.829687} {\bibfield
  {journal} {\bibinfo  {journal} {Contemporary Physics}\ }\textbf {\bibinfo
  {volume} {54}},\ \bibinfo {pages} {181} (\bibinfo {year} {2013})}\BibitemShut
  {NoStop}%
\bibitem [{\citenamefont {Oppenheim}\ \emph {et~al.}(2003)\citenamefont
  {Oppenheim}, \citenamefont {Horodecki}, \citenamefont {Horodecki},
  \citenamefont {Horodecki},\ and\ \citenamefont {Horodecki}}]{Oppenheim2003}%
  \BibitemOpen
  \bibfield  {author} {\bibinfo {author} {\bibfnamefont {J.}~\bibnamefont
  {Oppenheim}}, \bibinfo {author} {\bibfnamefont {K.}~\bibnamefont
  {Horodecki}}, \bibinfo {author} {\bibfnamefont {M.}~\bibnamefont
  {Horodecki}}, \bibinfo {author} {\bibfnamefont {P.}~\bibnamefont
  {Horodecki}},\ and\ \bibinfo {author} {\bibfnamefont {R.}~\bibnamefont
  {Horodecki}},\ }\bibfield  {title} {\bibinfo {title} {Mutually exclusive
  aspects of information carried by physical systems: Complementarity between
  local and nonlocal information},\ }\href
  {https://doi.org/10.1103/PhysRevA.68.022307} {\bibfield  {journal} {\bibinfo
  {journal} {Phys. Rev. A}\ }\textbf {\bibinfo {volume} {68}},\ \bibinfo
  {pages} {022307} (\bibinfo {year} {2003})}\BibitemShut {NoStop}%
\bibitem [{\citenamefont {Asb\'oth}\ \emph {et~al.}(2005)\citenamefont
  {Asb\'oth}, \citenamefont {Calsamiglia},\ and\ \citenamefont
  {Ritsch}}]{Asboth2005}%
  \BibitemOpen
  \bibfield  {author} {\bibinfo {author} {\bibfnamefont {J.~K.}\ \bibnamefont
  {Asb\'oth}}, \bibinfo {author} {\bibfnamefont {J.}~\bibnamefont
  {Calsamiglia}},\ and\ \bibinfo {author} {\bibfnamefont {H.}~\bibnamefont
  {Ritsch}},\ }\bibfield  {title} {\bibinfo {title} {Computable measure of
  nonclassicality for light},\ }\href
  {https://doi.org/10.1103/PhysRevLett.94.173602} {\bibfield  {journal}
  {\bibinfo  {journal} {Phys. Rev. Lett.}\ }\textbf {\bibinfo {volume} {94}},\
  \bibinfo {pages} {173602} (\bibinfo {year} {2005})}\BibitemShut {NoStop}%
\bibitem [{\citenamefont {Yao}\ \emph {et~al.}(2015)\citenamefont {Yao},
  \citenamefont {Xiao}, \citenamefont {Ge},\ and\ \citenamefont
  {Sun}}]{Yao2015}%
  \BibitemOpen
  \bibfield  {author} {\bibinfo {author} {\bibfnamefont {Y.}~\bibnamefont
  {Yao}}, \bibinfo {author} {\bibfnamefont {X.}~\bibnamefont {Xiao}}, \bibinfo
  {author} {\bibfnamefont {L.}~\bibnamefont {Ge}},\ and\ \bibinfo {author}
  {\bibfnamefont {C.~P.}\ \bibnamefont {Sun}},\ }\bibfield  {title} {\bibinfo
  {title} {Quantum coherence in multipartite systems},\ }\href
  {https://doi.org/10.1103/PhysRevA.92.022112} {\bibfield  {journal} {\bibinfo
  {journal} {Phys. Rev. A}\ }\textbf {\bibinfo {volume} {92}},\ \bibinfo
  {pages} {022112} (\bibinfo {year} {2015})}\BibitemShut {NoStop}%
\bibitem [{\citenamefont {Streltsov}\ \emph {et~al.}(2015)\citenamefont
  {Streltsov}, \citenamefont {Singh}, \citenamefont {Dhar}, \citenamefont
  {Bera},\ and\ \citenamefont {Adesso}}]{Streltsov2015}%
  \BibitemOpen
  \bibfield  {author} {\bibinfo {author} {\bibfnamefont {A.}~\bibnamefont
  {Streltsov}}, \bibinfo {author} {\bibfnamefont {U.}~\bibnamefont {Singh}},
  \bibinfo {author} {\bibfnamefont {H.~S.}\ \bibnamefont {Dhar}}, \bibinfo
  {author} {\bibfnamefont {M.~N.}\ \bibnamefont {Bera}},\ and\ \bibinfo
  {author} {\bibfnamefont {G.}~\bibnamefont {Adesso}},\ }\bibfield  {title}
  {\bibinfo {title} {Measuring quantum coherence with entanglement},\ }\href
  {https://doi.org/10.1103/PhysRevLett.115.020403} {\bibfield  {journal}
  {\bibinfo  {journal} {Phys. Rev. Lett.}\ }\textbf {\bibinfo {volume} {115}},\
  \bibinfo {pages} {020403} (\bibinfo {year} {2015})}\BibitemShut {NoStop}%
\bibitem [{\citenamefont {Xi}\ \emph {et~al.}(2015)\citenamefont {Xi},
  \citenamefont {Li},\ and\ \citenamefont {Fan}}]{Xi2015}%
  \BibitemOpen
  \bibfield  {author} {\bibinfo {author} {\bibfnamefont {Z.}~\bibnamefont
  {Xi}}, \bibinfo {author} {\bibfnamefont {Y.}~\bibnamefont {Li}},\ and\
  \bibinfo {author} {\bibfnamefont {H.}~\bibnamefont {Fan}},\ }\bibfield
  {title} {\bibinfo {title} {Measuring quantum coherence with entanglement},\
  }\href {https://doi.org/10.1038/srep10922} {\bibfield  {journal} {\bibinfo
  {journal} {Sci. Rep.}\ }\textbf {\bibinfo {volume} {5}},\ \bibinfo {pages}
  {10922} (\bibinfo {year} {2015})}\BibitemShut {NoStop}%
\bibitem [{\citenamefont {Chitambar}\ and\ \citenamefont
  {Hsieh}(2016)}]{Chitambar2016}%
  \BibitemOpen
  \bibfield  {author} {\bibinfo {author} {\bibfnamefont {E.}~\bibnamefont
  {Chitambar}}\ and\ \bibinfo {author} {\bibfnamefont {M.-H.}\ \bibnamefont
  {Hsieh}},\ }\bibfield  {title} {\bibinfo {title} {Relating the resource
  theories of entanglement and quantum coherence},\ }\href
  {https://doi.org/10.1103/PhysRevLett.117.020402} {\bibfield  {journal}
  {\bibinfo  {journal} {Phys. Rev. Lett.}\ }\textbf {\bibinfo {volume} {117}},\
  \bibinfo {pages} {020402} (\bibinfo {year} {2016})}\BibitemShut {NoStop}%
\bibitem [{\citenamefont {Killoran}\ \emph {et~al.}(2016)\citenamefont
  {Killoran}, \citenamefont {Steinhoff},\ and\ \citenamefont
  {Plenio}}]{Killoran2016}%
  \BibitemOpen
  \bibfield  {author} {\bibinfo {author} {\bibfnamefont {N.}~\bibnamefont
  {Killoran}}, \bibinfo {author} {\bibfnamefont {F.~E.~S.}\ \bibnamefont
  {Steinhoff}},\ and\ \bibinfo {author} {\bibfnamefont {M.~B.}\ \bibnamefont
  {Plenio}},\ }\bibfield  {title} {\bibinfo {title} {Converting nonclassicality
  into entanglement},\ }\href {https://doi.org/10.1103/PhysRevLett.116.080402}
  {\bibfield  {journal} {\bibinfo  {journal} {Phys. Rev. Lett.}\ }\textbf
  {\bibinfo {volume} {116}},\ \bibinfo {pages} {080402} (\bibinfo {year}
  {2016})}\BibitemShut {NoStop}%
\bibitem [{\citenamefont {Streltsov}\ \emph {et~al.}(2016)\citenamefont
  {Streltsov}, \citenamefont {Chitambar}, \citenamefont {Rana}, \citenamefont
  {Bera}, \citenamefont {Winter},\ and\ \citenamefont
  {Lewenstein}}]{Streltsov2016}%
  \BibitemOpen
  \bibfield  {author} {\bibinfo {author} {\bibfnamefont {A.}~\bibnamefont
  {Streltsov}}, \bibinfo {author} {\bibfnamefont {E.}~\bibnamefont
  {Chitambar}}, \bibinfo {author} {\bibfnamefont {S.}~\bibnamefont {Rana}},
  \bibinfo {author} {\bibfnamefont {M.~N.}\ \bibnamefont {Bera}}, \bibinfo
  {author} {\bibfnamefont {A.}~\bibnamefont {Winter}},\ and\ \bibinfo {author}
  {\bibfnamefont {M.}~\bibnamefont {Lewenstein}},\ }\bibfield  {title}
  {\bibinfo {title} {Entanglement and coherence in quantum state merging},\
  }\href {https://doi.org/10.1103/PhysRevLett.116.240405} {\bibfield  {journal}
  {\bibinfo  {journal} {Phys. Rev. Lett.}\ }\textbf {\bibinfo {volume} {116}},\
  \bibinfo {pages} {240405} (\bibinfo {year} {2016})}\BibitemShut {NoStop}%
\bibitem [{\citenamefont {Qi}\ \emph {et~al.}(2017)\citenamefont {Qi},
  \citenamefont {Gao},\ and\ \citenamefont {Yan}}]{Qi_2017}%
  \BibitemOpen
  \bibfield  {author} {\bibinfo {author} {\bibfnamefont {X.}~\bibnamefont
  {Qi}}, \bibinfo {author} {\bibfnamefont {T.}~\bibnamefont {Gao}},\ and\
  \bibinfo {author} {\bibfnamefont {F.}~\bibnamefont {Yan}},\ }\bibfield
  {title} {\bibinfo {title} {Measuring coherence with entanglement
  concurrence},\ }\href {https://doi.org/10.1088/1751-8121/aa7638} {\bibfield
  {journal} {\bibinfo  {journal} {Journal of Physics A: Mathematical and
  Theoretical}\ }\textbf {\bibinfo {volume} {50}},\ \bibinfo {pages} {285301}
  (\bibinfo {year} {2017})}\BibitemShut {NoStop}%
\bibitem [{\citenamefont {Zhu}\ \emph {et~al.}(2017)\citenamefont {Zhu},
  \citenamefont {Ma}, \citenamefont {Cao}, \citenamefont {Fei},\ and\
  \citenamefont {Vedral}}]{Zhu2017}%
  \BibitemOpen
  \bibfield  {author} {\bibinfo {author} {\bibfnamefont {H.}~\bibnamefont
  {Zhu}}, \bibinfo {author} {\bibfnamefont {Z.}~\bibnamefont {Ma}}, \bibinfo
  {author} {\bibfnamefont {Z.}~\bibnamefont {Cao}}, \bibinfo {author}
  {\bibfnamefont {S.-M.}\ \bibnamefont {Fei}},\ and\ \bibinfo {author}
  {\bibfnamefont {V.}~\bibnamefont {Vedral}},\ }\bibfield  {title} {\bibinfo
  {title} {Operational one-to-one mapping between coherence and entanglement
  measures},\ }\href {https://doi.org/10.1103/PhysRevA.96.032316} {\bibfield
  {journal} {\bibinfo  {journal} {Phys. Rev. A}\ }\textbf {\bibinfo {volume}
  {96}},\ \bibinfo {pages} {032316} (\bibinfo {year} {2017})}\BibitemShut
  {NoStop}%
\bibitem [{\citenamefont {Chin}(2017)}]{Chin2017}%
  \BibitemOpen
  \bibfield  {author} {\bibinfo {author} {\bibfnamefont {S.}~\bibnamefont
  {Chin}},\ }\bibfield  {title} {\bibinfo {title} {Coherence number as a
  discrete quantum resource},\ }\href
  {https://doi.org/10.1103/PhysRevA.96.042336} {\bibfield  {journal} {\bibinfo
  {journal} {Phys. Rev. A}\ }\textbf {\bibinfo {volume} {96}},\ \bibinfo
  {pages} {042336} (\bibinfo {year} {2017})}\BibitemShut {NoStop}%
\bibitem [{\citenamefont {Zhu}\ \emph {et~al.}(2018)\citenamefont {Zhu},
  \citenamefont {Hayashi},\ and\ \citenamefont {Chen}}]{Zhu2018}%
  \BibitemOpen
  \bibfield  {author} {\bibinfo {author} {\bibfnamefont {H.}~\bibnamefont
  {Zhu}}, \bibinfo {author} {\bibfnamefont {M.}~\bibnamefont {Hayashi}},\ and\
  \bibinfo {author} {\bibfnamefont {L.}~\bibnamefont {Chen}},\ }\bibfield
  {title} {\bibinfo {title} {Axiomatic and operational connections between the
  ${l}_{1}$-norm of coherence and negativity},\ }\href
  {https://doi.org/10.1103/PhysRevA.97.022342} {\bibfield  {journal} {\bibinfo
  {journal} {Phys. Rev. A}\ }\textbf {\bibinfo {volume} {97}},\ \bibinfo
  {pages} {022342} (\bibinfo {year} {2018})}\BibitemShut {NoStop}%
\bibitem [{\citenamefont {Egloff}\ \emph {et~al.}(2018)\citenamefont {Egloff},
  \citenamefont {Matera}, \citenamefont {Theurer},\ and\ \citenamefont
  {Plenio}}]{Egloff2018}%
  \BibitemOpen
  \bibfield  {author} {\bibinfo {author} {\bibfnamefont {D.}~\bibnamefont
  {Egloff}}, \bibinfo {author} {\bibfnamefont {J.~M.}\ \bibnamefont {Matera}},
  \bibinfo {author} {\bibfnamefont {T.}~\bibnamefont {Theurer}},\ and\ \bibinfo
  {author} {\bibfnamefont {M.~B.}\ \bibnamefont {Plenio}},\ }\bibfield  {title}
  {\bibinfo {title} {Of local operations and physical wires},\ }\href
  {https://doi.org/10.1103/PhysRevX.8.031005} {\bibfield  {journal} {\bibinfo
  {journal} {Phys. Rev. X}\ }\textbf {\bibinfo {volume} {8}},\ \bibinfo {pages}
  {031005} (\bibinfo {year} {2018})}\BibitemShut {NoStop}%
\bibitem [{\citenamefont {Streltsov}\ \emph {et~al.}(2018)\citenamefont
  {Streltsov}, \citenamefont {Kampermann}, \citenamefont {Wölk}, \citenamefont
  {Gessner},\ and\ \citenamefont {Bruß}}]{Streltsov_2018}%
  \BibitemOpen
  \bibfield  {author} {\bibinfo {author} {\bibfnamefont {A.}~\bibnamefont
  {Streltsov}}, \bibinfo {author} {\bibfnamefont {H.}~\bibnamefont
  {Kampermann}}, \bibinfo {author} {\bibfnamefont {S.}~\bibnamefont {Wölk}},
  \bibinfo {author} {\bibfnamefont {M.}~\bibnamefont {Gessner}},\ and\ \bibinfo
  {author} {\bibfnamefont {D.}~\bibnamefont {Bruß}},\ }\bibfield  {title}
  {\bibinfo {title} {Maximal coherence and the resource theory of purity},\
  }\href {https://doi.org/10.1088/1367-2630/aac484} {\bibfield  {journal}
  {\bibinfo  {journal} {New Journal of Physics}\ }\textbf {\bibinfo {volume}
  {20}},\ \bibinfo {pages} {053058} (\bibinfo {year} {2018})}\BibitemShut
  {NoStop}%
\bibitem [{\citenamefont {Kraemer}\ and\ \citenamefont {del
  Rio}(2021)}]{Kraemer2021}%
  \BibitemOpen
  \bibfield  {author} {\bibinfo {author} {\bibfnamefont {L.}~\bibnamefont
  {Kraemer}}\ and\ \bibinfo {author} {\bibfnamefont {L.}~\bibnamefont {del
  Rio}},\ }\bibfield  {title} {\bibinfo {title} {Currencies in resource
  theories},\ }\href {https://doi.org/10.3390/e23060755} {\bibfield  {journal}
  {\bibinfo  {journal} {Entropy}\ }\textbf {\bibinfo {volume} {23}},\ \bibinfo
  {pages} {755} (\bibinfo {year} {2021})}\BibitemShut {NoStop}%
\bibitem [{\citenamefont {Mekala}\ and\ \citenamefont
  {Sen}(2021)}]{relation_ent-coh}%
  \BibitemOpen
  \bibfield  {author} {\bibinfo {author} {\bibfnamefont {A.}~\bibnamefont
  {Mekala}}\ and\ \bibinfo {author} {\bibfnamefont {U.}~\bibnamefont {Sen}},\
  }\bibfield  {title} {\bibinfo {title} {All entangled states are quantum
  coherent with locally distinguishable bases},\ }\href
  {https://link.aps.org/doi/10.1103/PhysRevA.104.L050402} {\bibfield  {journal}
  {\bibinfo  {journal} {Phys. Rev. A}\ }\textbf {\bibinfo {volume} {104}},\
  \bibinfo {pages} {L050402} (\bibinfo {year} {2021})}\BibitemShut {NoStop}%
\bibitem [{\citenamefont {Kim}\ and\ \citenamefont {Lee}(2022)}]{Kim2022}%
  \BibitemOpen
  \bibfield  {author} {\bibinfo {author} {\bibfnamefont {H.-J.}\ \bibnamefont
  {Kim}}\ and\ \bibinfo {author} {\bibfnamefont {S.}~\bibnamefont {Lee}},\
  }\bibfield  {title} {\bibinfo {title} {Relation between quantum coherence and
  quantum entanglement in quantum measurements},\ }\href
  {https://doi.org/10.1103/PhysRevA.106.022401} {\bibfield  {journal} {\bibinfo
   {journal} {Phys. Rev. A}\ }\textbf {\bibinfo {volume} {106}},\ \bibinfo
  {pages} {022401} (\bibinfo {year} {2022})}\BibitemShut {NoStop}%
\bibitem [{\citenamefont {Liu}\ \emph {et~al.}(2022)\citenamefont {Liu},
  \citenamefont {Yang},\ and\ \citenamefont {Yan}}]{Liu2022}%
  \BibitemOpen
  \bibfield  {author} {\bibinfo {author} {\bibfnamefont {Y.}~\bibnamefont
  {Liu}}, \bibinfo {author} {\bibfnamefont {L.}~\bibnamefont {Yang}},\ and\
  \bibinfo {author} {\bibfnamefont {D.}~\bibnamefont {Yan}},\ }\bibfield
  {title} {\bibinfo {title} {The relation between entanglement measure and
  coherence measure based on {H}ellinger distance},\ }\href
  {https://doi.org/10.1007/s11128-022-03465-1} {\bibfield  {journal} {\bibinfo
  {journal} {Quantum Information Processing}\ }\textbf {\bibinfo {volume}
  {21}},\ \bibinfo {pages} {132} (\bibinfo {year} {2022})}\BibitemShut
  {NoStop}%
\bibitem [{\citenamefont {Bhattacharyya}\ \emph {et~al.}(2023)\citenamefont
  {Bhattacharyya}, \citenamefont {Ghoshal},\ and\ \citenamefont
  {Sen}}]{Bhattacharyya2021}%
  \BibitemOpen
  \bibfield  {author} {\bibinfo {author} {\bibfnamefont {A.}~\bibnamefont
  {Bhattacharyya}}, \bibinfo {author} {\bibfnamefont {A.}~\bibnamefont
  {Ghoshal}},\ and\ \bibinfo {author} {\bibfnamefont {U.}~\bibnamefont {Sen}},\
  }\bibfield  {title} {\bibinfo {title} {Correlation between
  resource-generating powers of quantum gates},\ }\href
  {https://doi.org/10.1103/PhysRevA.107.032406} {\bibfield  {journal} {\bibinfo
   {journal} {Phys. Rev. A}\ }\textbf {\bibinfo {volume} {107}},\ \bibinfo
  {pages} {032406} (\bibinfo {year} {2023})}\BibitemShut {NoStop}%
\bibitem [{\citenamefont {Zhang}\ \emph {et~al.}(2024)\citenamefont {Zhang},
  \citenamefont {Smith}, \citenamefont {Smolin}, \citenamefont {Liu},
  \citenamefont {Peng}, \citenamefont {Zhao}, \citenamefont {Girolami},
  \citenamefont {Ma}, \citenamefont {Yuan},\ and\ \citenamefont
  {Lu}}]{Zhang2024}%
  \BibitemOpen
  \bibfield  {author} {\bibinfo {author} {\bibfnamefont {T.}~\bibnamefont
  {Zhang}}, \bibinfo {author} {\bibfnamefont {G.}~\bibnamefont {Smith}},
  \bibinfo {author} {\bibfnamefont {J.~A.}\ \bibnamefont {Smolin}}, \bibinfo
  {author} {\bibfnamefont {L.}~\bibnamefont {Liu}}, \bibinfo {author}
  {\bibfnamefont {X.-J.}\ \bibnamefont {Peng}}, \bibinfo {author}
  {\bibfnamefont {Q.}~\bibnamefont {Zhao}}, \bibinfo {author} {\bibfnamefont
  {D.}~\bibnamefont {Girolami}}, \bibinfo {author} {\bibfnamefont
  {X.}~\bibnamefont {Ma}}, \bibinfo {author} {\bibfnamefont {X.}~\bibnamefont
  {Yuan}},\ and\ \bibinfo {author} {\bibfnamefont {H.}~\bibnamefont {Lu}},\
  }\bibfield  {title} {\bibinfo {title} {Quantification of entanglement and
  coherence with purity detection},\ }\href
  {https://doi.org/10.1038/s41534-024-00857-2} {\bibfield  {journal} {\bibinfo
  {journal} {npj Quantum Information}\ }\textbf {\bibinfo {volume} {10}},\
  \bibinfo {pages} {60} (\bibinfo {year} {2024})}\BibitemShut {NoStop}%
\bibitem [{\citenamefont {Laskowski}\ \emph {et~al.}(2010)\citenamefont
  {Laskowski}, \citenamefont {Paterek}, \citenamefont {\v{C}. Brukner},\ and\
  \citenamefont {\ifmmode~\dot{Z}\else \.{Z}\fi{}ukowski}}]{Laskowski2010}%
  \BibitemOpen
  \bibfield  {author} {\bibinfo {author} {\bibfnamefont {W.}~\bibnamefont
  {Laskowski}}, \bibinfo {author} {\bibfnamefont {T.}~\bibnamefont {Paterek}},
  \bibinfo {author} {\bibnamefont {\v{C}. Brukner}},\ and\ \bibinfo {author}
  {\bibfnamefont {M.}~\bibnamefont {\ifmmode~\dot{Z}\else \.{Z}\fi{}ukowski}},\
  }\bibfield  {title} {\bibinfo {title} {Entanglement and
  communication-reducing properties of noisy $n$-qubit states},\ }\href
  {https://doi.org/10.1103/PhysRevA.81.042101} {\bibfield  {journal} {\bibinfo
  {journal} {Phys. Rev. A}\ }\textbf {\bibinfo {volume} {81}},\ \bibinfo
  {pages} {042101} (\bibinfo {year} {2010})}\BibitemShut {NoStop}%
\bibitem [{\citenamefont {G\"{u}hne}\ and\ \citenamefont
  {Seevinck}(2010)}]{G_hne_2010}%
  \BibitemOpen
  \bibfield  {author} {\bibinfo {author} {\bibfnamefont {O.}~\bibnamefont
  {G\"{u}hne}}\ and\ \bibinfo {author} {\bibfnamefont {M.}~\bibnamefont
  {Seevinck}},\ }\bibfield  {title} {\bibinfo {title} {Separability criteria
  for genuine multiparticle entanglement},\ }\href
  {https://doi.org/10.1088/1367-2630/12/5/053002} {\bibfield  {journal}
  {\bibinfo  {journal} {New Journal of Physics}\ }\textbf {\bibinfo {volume}
  {12}},\ \bibinfo {pages} {053002} (\bibinfo {year} {2010})}\BibitemShut
  {NoStop}%
\bibitem [{\citenamefont {Huber}\ \emph {et~al.}(2010)\citenamefont {Huber},
  \citenamefont {Mintert}, \citenamefont {Gabriel},\ and\ \citenamefont
  {Hiesmayr}}]{Huber2010}%
  \BibitemOpen
  \bibfield  {author} {\bibinfo {author} {\bibfnamefont {M.}~\bibnamefont
  {Huber}}, \bibinfo {author} {\bibfnamefont {F.}~\bibnamefont {Mintert}},
  \bibinfo {author} {\bibfnamefont {A.}~\bibnamefont {Gabriel}},\ and\ \bibinfo
  {author} {\bibfnamefont {B.~C.}\ \bibnamefont {Hiesmayr}},\ }\bibfield
  {title} {\bibinfo {title} {Detection of high-dimensional genuine multipartite
  entanglement of mixed states},\ }\href
  {https://doi.org/10.1103/PhysRevLett.104.210501} {\bibfield  {journal}
  {\bibinfo  {journal} {Phys. Rev. Lett.}\ }\textbf {\bibinfo {volume} {104}},\
  \bibinfo {pages} {210501} (\bibinfo {year} {2010})}\BibitemShut {NoStop}%
\bibitem [{\citenamefont {Gabriel}\ \emph {et~al.}(2010)\citenamefont
  {Gabriel}, \citenamefont {Hiesmayr},\ and\ \citenamefont
  {Huber}}]{Gabriel2010}%
  \BibitemOpen
  \bibfield  {author} {\bibinfo {author} {\bibfnamefont {A.}~\bibnamefont
  {Gabriel}}, \bibinfo {author} {\bibfnamefont {B.~C.}\ \bibnamefont
  {Hiesmayr}},\ and\ \bibinfo {author} {\bibfnamefont {M.}~\bibnamefont
  {Huber}},\ }\bibfield  {title} {\bibinfo {title} {Criterion for
  $k$-separability in mixed multipartite systems},\ }\href@noop {} {\bibfield
  {journal} {\bibinfo  {journal} {Quantum Inf. Comput.}\ }\textbf {\bibinfo
  {volume} {10}},\ \bibinfo {pages} {829} (\bibinfo {year} {2010})}\BibitemShut
  {NoStop}%
\bibitem [{\citenamefont {Ananth}\ \emph {et~al.}(2015)\citenamefont {Ananth},
  \citenamefont {Chandrasekar},\ and\ \citenamefont
  {Senthilvelan}}]{Ananth2015}%
  \BibitemOpen
  \bibfield  {author} {\bibinfo {author} {\bibfnamefont {N.}~\bibnamefont
  {Ananth}}, \bibinfo {author} {\bibfnamefont {V.~K.}\ \bibnamefont
  {Chandrasekar}},\ and\ \bibinfo {author} {\bibfnamefont {M.}~\bibnamefont
  {Senthilvelan}},\ }\bibfield  {title} {\bibinfo {title} {Criteria for
  non-$k$-separability of $n$-partite quantum states},\ }\href
  {https://doi.org/10.1140/epjd/e2015-50538-5} {\bibfield  {journal} {\bibinfo
  {journal} {Eur. Phys. J. D}\ }\textbf {\bibinfo {volume} {69}},\ \bibinfo
  {pages} {56} (\bibinfo {year} {2015})}\BibitemShut {NoStop}%
\bibitem [{\citenamefont {Das}\ \emph {et~al.}()\citenamefont {Das},
  \citenamefont {Chanda}, \citenamefont {Lewenstein}, \citenamefont {Sanpera},
  \citenamefont {De},\ and\ \citenamefont {Sen}}]{Das2017}%
  \BibitemOpen
  \bibfield  {author} {\bibinfo {author} {\bibfnamefont {S.}~\bibnamefont
  {Das}}, \bibinfo {author} {\bibfnamefont {T.}~\bibnamefont {Chanda}},
  \bibinfo {author} {\bibfnamefont {M.}~\bibnamefont {Lewenstein}}, \bibinfo
  {author} {\bibfnamefont {A.}~\bibnamefont {Sanpera}}, \bibinfo {author}
  {\bibfnamefont {A.~S.}\ \bibnamefont {De}},\ and\ \bibinfo {author}
  {\bibfnamefont {U.}~\bibnamefont {Sen}},\ }\bibfield  {title} {\bibinfo
  {title} {The separability versus entanglement problem},\ }\href@noop {}
  {\bibinfo  {journal} {arXiv:1701.02187}\ }\BibitemShut {NoStop}%
\bibitem [{\citenamefont {Gao}\ and\ \citenamefont {Hong}(2010)}]{Gao2010}%
  \BibitemOpen
\bibfield  {journal} {  }\bibfield  {author} {\bibinfo {author} {\bibfnamefont
  {T.}~\bibnamefont {Gao}}\ and\ \bibinfo {author} {\bibfnamefont
  {Y.}~\bibnamefont {Hong}},\ }\bibfield  {title} {\bibinfo {title} {Detection
  of genuinely entangled and nonseparable $n$-partite quantum states},\ }\href
  {https://doi.org/10.1103/PhysRevA.82.062113} {\bibfield  {journal} {\bibinfo
  {journal} {Phys. Rev. A}\ }\textbf {\bibinfo {volume} {82}},\ \bibinfo
  {pages} {062113} (\bibinfo {year} {2010})}\BibitemShut {NoStop}%
\bibitem [{\citenamefont {Gao}\ and\ \citenamefont {Hong}(2011)}]{Gao2011}%
  \BibitemOpen
  \bibfield  {author} {\bibinfo {author} {\bibfnamefont {T.}~\bibnamefont
  {Gao}}\ and\ \bibinfo {author} {\bibfnamefont {Y.}~\bibnamefont {Hong}},\
  }\bibfield  {title} {\bibinfo {title} {Separability criteria for several
  classes of $n$-partite quantum states},\ }\href
  {https://doi.org/10.1140/epjd/e2010-10432-4} {\bibfield  {journal} {\bibinfo
  {journal} {Eur. Phys. J. D}\ }\textbf {\bibinfo {volume} {61}},\ \bibinfo
  {pages} {765} (\bibinfo {year} {2011})}\BibitemShut {NoStop}%
\bibitem [{\citenamefont {Wu}\ \emph {et~al.}(2012)\citenamefont {Wu},
  \citenamefont {Kampermann}, \citenamefont {Bru\ss{}}, \citenamefont
  {Kl\"ockl},\ and\ \citenamefont {Huber}}]{Wu2012}%
  \BibitemOpen
  \bibfield  {author} {\bibinfo {author} {\bibfnamefont {J.-Y.}\ \bibnamefont
  {Wu}}, \bibinfo {author} {\bibfnamefont {H.}~\bibnamefont {Kampermann}},
  \bibinfo {author} {\bibfnamefont {D.}~\bibnamefont {Bru\ss{}}}, \bibinfo
  {author} {\bibfnamefont {C.}~\bibnamefont {Kl\"ockl}},\ and\ \bibinfo
  {author} {\bibfnamefont {M.}~\bibnamefont {Huber}},\ }\bibfield  {title}
  {\bibinfo {title} {Determining lower bounds on a measure of multipartite
  entanglement from few local observables},\ }\href
  {https://doi.org/10.1103/PhysRevA.86.022319} {\bibfield  {journal} {\bibinfo
  {journal} {Phys. Rev. A}\ }\textbf {\bibinfo {volume} {86}},\ \bibinfo
  {pages} {022319} (\bibinfo {year} {2012})}\BibitemShut {NoStop}%
\bibitem [{\citenamefont {Chen}\ \emph {et~al.}(2012)\citenamefont {Chen},
  \citenamefont {Ma}, \citenamefont {Chen},\ and\ \citenamefont
  {Severini}}]{Chen2012}%
  \BibitemOpen
  \bibfield  {author} {\bibinfo {author} {\bibfnamefont {Z.-H.}\ \bibnamefont
  {Chen}}, \bibinfo {author} {\bibfnamefont {Z.-H.}\ \bibnamefont {Ma}},
  \bibinfo {author} {\bibfnamefont {J.-L.}\ \bibnamefont {Chen}},\ and\
  \bibinfo {author} {\bibfnamefont {S.}~\bibnamefont {Severini}},\ }\bibfield
  {title} {\bibinfo {title} {Improved lower bounds on
  genuine-multipartite-entanglement concurrence},\ }\href
  {https://doi.org/10.1103/PhysRevA.85.062320} {\bibfield  {journal} {\bibinfo
  {journal} {Phys. Rev. A}\ }\textbf {\bibinfo {volume} {85}},\ \bibinfo
  {pages} {062320} (\bibinfo {year} {2012})}\BibitemShut {NoStop}%
\bibitem [{\citenamefont {Vitagliano}\ \emph {et~al.}(2011)\citenamefont
  {Vitagliano}, \citenamefont {Hyllus}, \citenamefont {Egusquiza},\ and\
  \citenamefont {T\'oth}}]{Vitagliano2011}%
  \BibitemOpen
  \bibfield  {author} {\bibinfo {author} {\bibfnamefont {G.}~\bibnamefont
  {Vitagliano}}, \bibinfo {author} {\bibfnamefont {P.}~\bibnamefont {Hyllus}},
  \bibinfo {author} {\bibfnamefont {I.~L.}\ \bibnamefont {Egusquiza}},\ and\
  \bibinfo {author} {\bibfnamefont {G.}~\bibnamefont {T\'oth}},\ }\bibfield
  {title} {\bibinfo {title} {Spin squeezing inequalities for arbitrary spin},\
  }\href {https://doi.org/10.1103/PhysRevLett.107.240502} {\bibfield  {journal}
  {\bibinfo  {journal} {Phys. Rev. Lett.}\ }\textbf {\bibinfo {volume} {107}},\
  \bibinfo {pages} {240502} (\bibinfo {year} {2011})}\BibitemShut {NoStop}%
\bibitem [{\citenamefont {Seevinck}\ and\ \citenamefont
  {Svetlichny}(2002)}]{Seevinck2002}%
  \BibitemOpen
  \bibfield  {author} {\bibinfo {author} {\bibfnamefont {M.}~\bibnamefont
  {Seevinck}}\ and\ \bibinfo {author} {\bibfnamefont {G.}~\bibnamefont
  {Svetlichny}},\ }\bibfield  {title} {\bibinfo {title} {Bell-type inequalities
  for partial separability in $n$-particle systems and quantum mechanical
  violations},\ }\href {https://doi.org/10.1103/PhysRevLett.89.060401}
  {\bibfield  {journal} {\bibinfo  {journal} {Phys. Rev. Lett.}\ }\textbf
  {\bibinfo {volume} {89}},\ \bibinfo {pages} {060401} (\bibinfo {year}
  {2002})}\BibitemShut {NoStop}%
\bibitem [{\citenamefont {Doherty}\ \emph {et~al.}(2002)\citenamefont
  {Doherty}, \citenamefont {Parrilo},\ and\ \citenamefont
  {Spedalieri}}]{Doherty2002}%
  \BibitemOpen
  \bibfield  {author} {\bibinfo {author} {\bibfnamefont {A.~C.}\ \bibnamefont
  {Doherty}}, \bibinfo {author} {\bibfnamefont {P.~A.}\ \bibnamefont
  {Parrilo}},\ and\ \bibinfo {author} {\bibfnamefont {F.~M.}\ \bibnamefont
  {Spedalieri}},\ }\bibfield  {title} {\bibinfo {title} {Distinguishing
  separable and entangled states},\ }\href
  {https://doi.org/10.1103/PhysRevLett.88.187904} {\bibfield  {journal}
  {\bibinfo  {journal} {Phys. Rev. Lett.}\ }\textbf {\bibinfo {volume} {88}},\
  \bibinfo {pages} {187904} (\bibinfo {year} {2002})}\BibitemShut {NoStop}%
\bibitem [{\citenamefont {Doherty}\ \emph {et~al.}(2004)\citenamefont
  {Doherty}, \citenamefont {Parrilo},\ and\ \citenamefont
  {Spedalieri}}]{Doherty2004}%
  \BibitemOpen
  \bibfield  {author} {\bibinfo {author} {\bibfnamefont {A.~C.}\ \bibnamefont
  {Doherty}}, \bibinfo {author} {\bibfnamefont {P.~A.}\ \bibnamefont
  {Parrilo}},\ and\ \bibinfo {author} {\bibfnamefont {F.~M.}\ \bibnamefont
  {Spedalieri}},\ }\bibfield  {title} {\bibinfo {title} {Complete family of
  separability criteria},\ }\href {https://doi.org/10.1103/PhysRevA.69.022308}
  {\bibfield  {journal} {\bibinfo  {journal} {Phys. Rev. A}\ }\textbf {\bibinfo
  {volume} {69}},\ \bibinfo {pages} {022308} (\bibinfo {year}
  {2004})}\BibitemShut {NoStop}%
\bibitem [{\citenamefont {Doherty}\ \emph {et~al.}(2005)\citenamefont
  {Doherty}, \citenamefont {Parrilo},\ and\ \citenamefont
  {Spedalieri}}]{Doherty2005}%
  \BibitemOpen
  \bibfield  {author} {\bibinfo {author} {\bibfnamefont {A.~C.}\ \bibnamefont
  {Doherty}}, \bibinfo {author} {\bibfnamefont {P.~A.}\ \bibnamefont
  {Parrilo}},\ and\ \bibinfo {author} {\bibfnamefont {F.~M.}\ \bibnamefont
  {Spedalieri}},\ }\bibfield  {title} {\bibinfo {title} {Detecting multipartite
  entanglement},\ }\href {https://doi.org/10.1103/PhysRevA.71.032333}
  {\bibfield  {journal} {\bibinfo  {journal} {Phys. Rev. A}\ }\textbf {\bibinfo
  {volume} {71}},\ \bibinfo {pages} {032333} (\bibinfo {year}
  {2005})}\BibitemShut {NoStop}%
\bibitem [{\citenamefont {Gittsovich}\ \emph {et~al.}(2010)\citenamefont
  {Gittsovich}, \citenamefont {Hyllus},\ and\ \citenamefont
  {G\"uhne}}]{Gittsovich2010}%
  \BibitemOpen
  \bibfield  {author} {\bibinfo {author} {\bibfnamefont {O.}~\bibnamefont
  {Gittsovich}}, \bibinfo {author} {\bibfnamefont {P.}~\bibnamefont {Hyllus}},\
  and\ \bibinfo {author} {\bibfnamefont {O.}~\bibnamefont {G\"uhne}},\
  }\bibfield  {title} {\bibinfo {title} {Multiparticle covariance matrices and
  the impossibility of detecting graph-state entanglement with two-particle
  correlations},\ }\href {https://doi.org/10.1103/PhysRevA.82.032306}
  {\bibfield  {journal} {\bibinfo  {journal} {Phys. Rev. A}\ }\textbf {\bibinfo
  {volume} {82}},\ \bibinfo {pages} {032306} (\bibinfo {year}
  {2010})}\BibitemShut {NoStop}%
\bibitem [{\citenamefont {Hyllus}\ \emph {et~al.}(2012)\citenamefont {Hyllus},
  \citenamefont {Laskowski}, \citenamefont {Krischek}, \citenamefont
  {Schwemmer}, \citenamefont {Wieczorek}, \citenamefont {Weinfurter},
  \citenamefont {Pezz\'e},\ and\ \citenamefont {Smerzi}}]{Hyllus2012}%
  \BibitemOpen
  \bibfield  {author} {\bibinfo {author} {\bibfnamefont {P.}~\bibnamefont
  {Hyllus}}, \bibinfo {author} {\bibfnamefont {W.}~\bibnamefont {Laskowski}},
  \bibinfo {author} {\bibfnamefont {R.}~\bibnamefont {Krischek}}, \bibinfo
  {author} {\bibfnamefont {C.}~\bibnamefont {Schwemmer}}, \bibinfo {author}
  {\bibfnamefont {W.}~\bibnamefont {Wieczorek}}, \bibinfo {author}
  {\bibfnamefont {H.}~\bibnamefont {Weinfurter}}, \bibinfo {author}
  {\bibfnamefont {L.}~\bibnamefont {Pezz\'e}},\ and\ \bibinfo {author}
  {\bibfnamefont {A.}~\bibnamefont {Smerzi}},\ }\bibfield  {title} {\bibinfo
  {title} {Fisher information and multiparticle entanglement},\ }\href
  {https://doi.org/10.1103/PhysRevA.85.022321} {\bibfield  {journal} {\bibinfo
  {journal} {Phys. Rev. A}\ }\textbf {\bibinfo {volume} {85}},\ \bibinfo
  {pages} {022321} (\bibinfo {year} {2012})}\BibitemShut {NoStop}%
\bibitem [{\citenamefont {T\'oth}(2012)}]{Toth2012}%
  \BibitemOpen
  \bibfield  {author} {\bibinfo {author} {\bibfnamefont {G.}~\bibnamefont
  {T\'oth}},\ }\bibfield  {title} {\bibinfo {title} {Multipartite entanglement
  and high-precision metrology},\ }\href
  {https://doi.org/10.1103/PhysRevA.85.022322} {\bibfield  {journal} {\bibinfo
  {journal} {Phys. Rev. A}\ }\textbf {\bibinfo {volume} {85}},\ \bibinfo
  {pages} {022322} (\bibinfo {year} {2012})}\BibitemShut {NoStop}%
\bibitem [{\citenamefont {Hong}\ \emph {et~al.}(2012)\citenamefont {Hong},
  \citenamefont {Gao},\ and\ \citenamefont {Yan}}]{Hong2012}%
  \BibitemOpen
  \bibfield  {author} {\bibinfo {author} {\bibfnamefont {Y.}~\bibnamefont
  {Hong}}, \bibinfo {author} {\bibfnamefont {T.}~\bibnamefont {Gao}},\ and\
  \bibinfo {author} {\bibfnamefont {F.}~\bibnamefont {Yan}},\ }\bibfield
  {title} {\bibinfo {title} {Measure of multipartite entanglement with
  computable lower bounds},\ }\href
  {https://doi.org/10.1103/PhysRevA.86.062323} {\bibfield  {journal} {\bibinfo
  {journal} {Phys. Rev. A}\ }\textbf {\bibinfo {volume} {86}},\ \bibinfo
  {pages} {062323} (\bibinfo {year} {2012})}\BibitemShut {NoStop}%
\bibitem [{\citenamefont {Huber}\ \emph {et~al.}(2013)\citenamefont {Huber},
  \citenamefont {Perarnau-Llobet},\ and\ \citenamefont
  {de~Vicente}}]{Huber2013}%
  \BibitemOpen
  \bibfield  {author} {\bibinfo {author} {\bibfnamefont {M.}~\bibnamefont
  {Huber}}, \bibinfo {author} {\bibfnamefont {M.}~\bibnamefont
  {Perarnau-Llobet}},\ and\ \bibinfo {author} {\bibfnamefont {J.~I.}\
  \bibnamefont {de~Vicente}},\ }\bibfield  {title} {\bibinfo {title} {Entropy
  vector formalism and the structure of multidimensional entanglement in
  multipartite systems},\ }\href {https://doi.org/10.1103/PhysRevA.88.042328}
  {\bibfield  {journal} {\bibinfo  {journal} {Phys. Rev. A}\ }\textbf {\bibinfo
  {volume} {88}},\ \bibinfo {pages} {042328} (\bibinfo {year}
  {2013})}\BibitemShut {NoStop}%
\bibitem [{\citenamefont {Gao}\ \emph {et~al.}(2013)\citenamefont {Gao},
  \citenamefont {Hong}, \citenamefont {Lu},\ and\ \citenamefont
  {Yan}}]{Gao_2013}%
  \BibitemOpen
  \bibfield  {author} {\bibinfo {author} {\bibfnamefont {T.}~\bibnamefont
  {Gao}}, \bibinfo {author} {\bibfnamefont {Y.}~\bibnamefont {Hong}}, \bibinfo
  {author} {\bibfnamefont {Y.}~\bibnamefont {Lu}},\ and\ \bibinfo {author}
  {\bibfnamefont {F.}~\bibnamefont {Yan}},\ }\bibfield  {title} {\bibinfo
  {title} {Efficient k-separability criteria for mixed multipartite quantum
  states},\ }\href {https://doi.org/10.1209/0295-5075/104/20007} {\bibfield
  {journal} {\bibinfo  {journal} {Europhysics Letters}\ }\textbf {\bibinfo
  {volume} {104}},\ \bibinfo {pages} {20007} (\bibinfo {year}
  {2013})}\BibitemShut {NoStop}%
\bibitem [{\citenamefont {Gao}\ \emph {et~al.}(2014)\citenamefont {Gao},
  \citenamefont {Yan},\ and\ \citenamefont {van Enk}}]{Gao2014}%
  \BibitemOpen
  \bibfield  {author} {\bibinfo {author} {\bibfnamefont {T.}~\bibnamefont
  {Gao}}, \bibinfo {author} {\bibfnamefont {F.}~\bibnamefont {Yan}},\ and\
  \bibinfo {author} {\bibfnamefont {S.~J.}\ \bibnamefont {van Enk}},\
  }\bibfield  {title} {\bibinfo {title} {Permutationally invariant part of a
  density matrix and nonseparability of $n$-qubit states},\ }\href
  {https://doi.org/10.1103/PhysRevLett.112.180501} {\bibfield  {journal}
  {\bibinfo  {journal} {Phys. Rev. Lett.}\ }\textbf {\bibinfo {volume} {112}},\
  \bibinfo {pages} {180501} (\bibinfo {year} {2014})}\BibitemShut {NoStop}%
\bibitem [{\citenamefont {Kl\"ockl}\ and\ \citenamefont
  {Huber}(2015)}]{Klockl2015}%
  \BibitemOpen
  \bibfield  {author} {\bibinfo {author} {\bibfnamefont {C.}~\bibnamefont
  {Kl\"ockl}}\ and\ \bibinfo {author} {\bibfnamefont {M.}~\bibnamefont
  {Huber}},\ }\bibfield  {title} {\bibinfo {title} {Characterizing multipartite
  entanglement without shared reference frames},\ }\href
  {https://doi.org/10.1103/PhysRevA.91.042339} {\bibfield  {journal} {\bibinfo
  {journal} {Phys. Rev. A}\ }\textbf {\bibinfo {volume} {91}},\ \bibinfo
  {pages} {042339} (\bibinfo {year} {2015})}\BibitemShut {NoStop}%
\bibitem [{\citenamefont {Hong}\ \emph {et~al.}(2015)\citenamefont {Hong},
  \citenamefont {Luo},\ and\ \citenamefont {Song}}]{Hong2015}%
  \BibitemOpen
  \bibfield  {author} {\bibinfo {author} {\bibfnamefont {Y.}~\bibnamefont
  {Hong}}, \bibinfo {author} {\bibfnamefont {S.}~\bibnamefont {Luo}},\ and\
  \bibinfo {author} {\bibfnamefont {H.}~\bibnamefont {Song}},\ }\bibfield
  {title} {\bibinfo {title} {Detecting $k$-nonseparability via quantum fisher
  information},\ }\href {https://doi.org/10.1103/PhysRevA.91.042313} {\bibfield
   {journal} {\bibinfo  {journal} {Phys. Rev. A}\ }\textbf {\bibinfo {volume}
  {91}},\ \bibinfo {pages} {042313} (\bibinfo {year} {2015})}\BibitemShut
  {NoStop}%
\bibitem [{\citenamefont {Gessner}\ \emph {et~al.}(2016)\citenamefont
  {Gessner}, \citenamefont {Pezz\`e},\ and\ \citenamefont
  {Smerzi}}]{Gessner2016}%
  \BibitemOpen
  \bibfield  {author} {\bibinfo {author} {\bibfnamefont {M.}~\bibnamefont
  {Gessner}}, \bibinfo {author} {\bibfnamefont {L.}~\bibnamefont {Pezz\`e}},\
  and\ \bibinfo {author} {\bibfnamefont {A.}~\bibnamefont {Smerzi}},\
  }\bibfield  {title} {\bibinfo {title} {Efficient entanglement criteria for
  discrete, continuous, and hybrid variables},\ }\href
  {https://doi.org/10.1103/PhysRevA.94.020101} {\bibfield  {journal} {\bibinfo
  {journal} {Phys. Rev. A}\ }\textbf {\bibinfo {volume} {94}},\ \bibinfo
  {pages} {020101} (\bibinfo {year} {2016})}\BibitemShut {NoStop}%
\bibitem [{\citenamefont {Hong}\ and\ \citenamefont {Luo}(2016)}]{Hong2016}%
  \BibitemOpen
  \bibfield  {author} {\bibinfo {author} {\bibfnamefont {Y.}~\bibnamefont
  {Hong}}\ and\ \bibinfo {author} {\bibfnamefont {S.}~\bibnamefont {Luo}},\
  }\bibfield  {title} {\bibinfo {title} {Detecting $k$-nonseparability via
  local uncertainty relations},\ }\href
  {https://doi.org/10.1103/PhysRevA.93.042310} {\bibfield  {journal} {\bibinfo
  {journal} {Phys. Rev. A}\ }\textbf {\bibinfo {volume} {93}},\ \bibinfo
  {pages} {042310} (\bibinfo {year} {2016})}\BibitemShut {NoStop}%
\bibitem [{\citenamefont {Hong}\ \emph {et~al.}(2021)\citenamefont {Hong},
  \citenamefont {Gao},\ and\ \citenamefont {Yan}}]{HONG2021}%
  \BibitemOpen
  \bibfield  {author} {\bibinfo {author} {\bibfnamefont {Y.}~\bibnamefont
  {Hong}}, \bibinfo {author} {\bibfnamefont {T.}~\bibnamefont {Gao}},\ and\
  \bibinfo {author} {\bibfnamefont {F.}~\bibnamefont {Yan}},\ }\bibfield
  {title} {\bibinfo {title} {Detection of k-partite entanglement and
  k-nonseparability of multipartite quantum states},\ }\href
  {https://doi.org/https://doi.org/10.1016/j.physleta.2021.127347} {\bibfield
  {journal} {\bibinfo  {journal} {Physics Letters A}\ }\textbf {\bibinfo
  {volume} {401}},\ \bibinfo {pages} {127347} (\bibinfo {year}
  {2021})}\BibitemShut {NoStop}%
\bibitem [{\citenamefont {Szalay}(2019)}]{Szalay2019}%
  \BibitemOpen
  \bibfield  {author} {\bibinfo {author} {\bibfnamefont {S.}~\bibnamefont
  {Szalay}},\ }\bibfield  {title} {\bibinfo {title} {k-stretchability of
  entanglement, and the duality of k-separability and k-producibility},\ }\href
  {https://doi.org/10.22331/q-2019-12-02-204} {\bibfield  {journal} {\bibinfo
  {journal} {{Quantum}}\ }\textbf {\bibinfo {volume} {3}},\ \bibinfo {pages}
  {204} (\bibinfo {year} {2019})}\BibitemShut {NoStop}%
\bibitem [{\citenamefont {T{\'{o}}th}(2020)}]{Toth2020}%
  \BibitemOpen
  \bibfield  {author} {\bibinfo {author} {\bibfnamefont {G.}~\bibnamefont
  {T{\'{o}}th}},\ }\bibfield  {title} {\bibinfo {title} {Stretching the limits
  of multiparticle entanglement},\ }\href
  {https://doi.org/10.22331/qv-2020-01-27-30} {\bibfield  {journal} {\bibinfo
  {journal} {{Quantum Views}}\ }\textbf {\bibinfo {volume} {4}},\ \bibinfo
  {pages} {30} (\bibinfo {year} {2020})}\BibitemShut {NoStop}%
\bibitem [{\citenamefont {Ren}\ \emph {et~al.}(2021)\citenamefont {Ren},
  \citenamefont {Li}, \citenamefont {Smerzi},\ and\ \citenamefont
  {Gessner}}]{Ren2021}%
  \BibitemOpen
  \bibfield  {author} {\bibinfo {author} {\bibfnamefont {Z.}~\bibnamefont
  {Ren}}, \bibinfo {author} {\bibfnamefont {W.}~\bibnamefont {Li}}, \bibinfo
  {author} {\bibfnamefont {A.}~\bibnamefont {Smerzi}},\ and\ \bibinfo {author}
  {\bibfnamefont {M.}~\bibnamefont {Gessner}},\ }\bibfield  {title} {\bibinfo
  {title} {Metrological detection of multipartite entanglement from young
  diagrams},\ }\href {https://doi.org/10.1103/PhysRevLett.126.080502}
  {\bibfield  {journal} {\bibinfo  {journal} {Phys. Rev. Lett.}\ }\textbf
  {\bibinfo {volume} {126}},\ \bibinfo {pages} {080502} (\bibinfo {year}
  {2021})}\BibitemShut {NoStop}%
\bibitem [{\citenamefont {Bennett}\ \emph {et~al.}(1996)\citenamefont
  {Bennett}, \citenamefont {DiVincenzo}, \citenamefont {Smolin},\ and\
  \citenamefont {Wootters}}]{LOCC1}%
  \BibitemOpen
  \bibfield  {author} {\bibinfo {author} {\bibfnamefont {C.~H.}\ \bibnamefont
  {Bennett}}, \bibinfo {author} {\bibfnamefont {D.~P.}\ \bibnamefont
  {DiVincenzo}}, \bibinfo {author} {\bibfnamefont {J.~A.}\ \bibnamefont
  {Smolin}},\ and\ \bibinfo {author} {\bibfnamefont {W.}~\bibnamefont
  {Wootters}},\ }\bibfield  {title} {\bibinfo {title} {Mixed-state entanglement
  and quantum error correction},\ }\href
  {https://link.aps.org/doi/10.1103/PhysRevA.54.3824} {\bibfield  {journal}
  {\bibinfo  {journal} {Phys. Rev. A}\ }\textbf {\bibinfo {volume} {54}},\
  \bibinfo {pages} {3824} (\bibinfo {year} {1996})}\BibitemShut {NoStop}%
\bibitem [{\citenamefont {Vedral}\ and\ \citenamefont {Plenio}(1998)}]{LOCC2}%
  \BibitemOpen
  \bibfield  {author} {\bibinfo {author} {\bibfnamefont {V.}~\bibnamefont
  {Vedral}}\ and\ \bibinfo {author} {\bibfnamefont {M.~B.}\ \bibnamefont
  {Plenio}},\ }\bibfield  {title} {\bibinfo {title} {Entanglement measures and
  purification procedures},\ }\href
  {https://link.aps.org/doi/10.1103/PhysRevA.57.1619} {\bibfield  {journal}
  {\bibinfo  {journal} {Phys. Rev. A}\ }\textbf {\bibinfo {volume} {57}},\
  \bibinfo {pages} {1619} (\bibinfo {year} {1998})}\BibitemShut {NoStop}%
\bibitem [{\citenamefont {Rains}()}]{Rains1998}%
  \BibitemOpen
  \bibfield  {author} {\bibinfo {author} {\bibfnamefont {E.~M.}\ \bibnamefont
  {Rains}},\ }\bibfield  {title} {\bibinfo {title} {Entanglement purification
  via separable superoperators},\ }\href@noop {} {\bibinfo  {journal}
  {arXiv:quant-ph/9707002}\ }\BibitemShut {NoStop}%
\bibitem [{\citenamefont {Chitambar}\ \emph {et~al.}(2014)\citenamefont
  {Chitambar}, \citenamefont {Leung}, \citenamefont {Man{\v{c}}inska},
  \citenamefont {Ozols},\ and\ \citenamefont {Winter}}]{LOCC3}%
  \BibitemOpen
\bibfield  {journal} {  }\bibfield  {author} {\bibinfo {author} {\bibfnamefont
  {E.}~\bibnamefont {Chitambar}}, \bibinfo {author} {\bibfnamefont
  {D.}~\bibnamefont {Leung}}, \bibinfo {author} {\bibfnamefont
  {L.}~\bibnamefont {Man{\v{c}}inska}}, \bibinfo {author} {\bibfnamefont
  {M.}~\bibnamefont {Ozols}},\ and\ \bibinfo {author} {\bibfnamefont
  {A.}~\bibnamefont {Winter}},\ }\bibfield  {title} {\bibinfo {title}
  {Everything you always wanted to know about {LOCC} (but were afraid to
  ask)},\ }\href {https://doi.org/10.1007%2Fs00220-014-1953-9} {\bibfield
  {journal} {\bibinfo  {journal} {Communications in Mathematical Physics}\
  }\textbf {\bibinfo {volume} {328}},\ \bibinfo {pages} {303} (\bibinfo {year}
  {2014})}\BibitemShut {NoStop}%
\bibitem [{\citenamefont {Horodecki}\ \emph
  {et~al.}(2003{\natexlab{a}})\citenamefont {Horodecki}, \citenamefont
  {Sen(De)}, \citenamefont {Sen},\ and\ \citenamefont
  {Horodecki}}]{Horodecki2003}%
  \BibitemOpen
  \bibfield  {author} {\bibinfo {author} {\bibfnamefont {M.}~\bibnamefont
  {Horodecki}}, \bibinfo {author} {\bibfnamefont {A.}~\bibnamefont {Sen(De)}},
  \bibinfo {author} {\bibfnamefont {U.}~\bibnamefont {Sen}},\ and\ \bibinfo
  {author} {\bibfnamefont {K.}~\bibnamefont {Horodecki}},\ }\bibfield  {title}
  {\bibinfo {title} {Local indistinguishability: More nonlocality with less
  entanglement},\ }\href {https://doi.org/10.1103/PhysRevLett.90.047902}
  {\bibfield  {journal} {\bibinfo  {journal} {Phys. Rev. Lett.}\ }\textbf
  {\bibinfo {volume} {90}},\ \bibinfo {pages} {047902} (\bibinfo {year}
  {2003}{\natexlab{a}})}\BibitemShut {NoStop}%
\bibitem [{\citenamefont {Bennett}\ \emph {et~al.}(1999)\citenamefont
  {Bennett}, \citenamefont {DiVincenzo}, \citenamefont {Fuchs}, \citenamefont
  {Mor}, \citenamefont {Rains}, \citenamefont {Shor}, \citenamefont {Smolin},\
  and\ \citenamefont {Wootters}}]{Bennett1999}%
  \BibitemOpen
  \bibfield  {author} {\bibinfo {author} {\bibfnamefont {C.~H.}\ \bibnamefont
  {Bennett}}, \bibinfo {author} {\bibfnamefont {D.~P.}\ \bibnamefont
  {DiVincenzo}}, \bibinfo {author} {\bibfnamefont {C.~A.}\ \bibnamefont
  {Fuchs}}, \bibinfo {author} {\bibfnamefont {T.}~\bibnamefont {Mor}}, \bibinfo
  {author} {\bibfnamefont {E.}~\bibnamefont {Rains}}, \bibinfo {author}
  {\bibfnamefont {P.~W.}\ \bibnamefont {Shor}}, \bibinfo {author}
  {\bibfnamefont {J.~A.}\ \bibnamefont {Smolin}},\ and\ \bibinfo {author}
  {\bibfnamefont {W.~K.}\ \bibnamefont {Wootters}},\ }\bibfield  {title}
  {\bibinfo {title} {Quantum nonlocality without entanglement},\ }\href
  {https://doi.org/10.1103/PhysRevA.59.1070} {\bibfield  {journal} {\bibinfo
  {journal} {Phys. Rev. A}\ }\textbf {\bibinfo {volume} {59}},\ \bibinfo
  {pages} {1070} (\bibinfo {year} {1999})}\BibitemShut {NoStop}%
\bibitem [{\citenamefont {Walgate}\ \emph {et~al.}(2000)\citenamefont
  {Walgate}, \citenamefont {Short}, \citenamefont {Hardy},\ and\ \citenamefont
  {Vedral}}]{Walgate2000}%
  \BibitemOpen
  \bibfield  {author} {\bibinfo {author} {\bibfnamefont {J.}~\bibnamefont
  {Walgate}}, \bibinfo {author} {\bibfnamefont {A.~J.}\ \bibnamefont {Short}},
  \bibinfo {author} {\bibfnamefont {L.}~\bibnamefont {Hardy}},\ and\ \bibinfo
  {author} {\bibfnamefont {V.}~\bibnamefont {Vedral}},\ }\bibfield  {title}
  {\bibinfo {title} {Local distinguishability of multipartite orthogonal
  quantum states},\ }\href {https://doi.org/10.1103/PhysRevLett.85.4972}
  {\bibfield  {journal} {\bibinfo  {journal} {Phys. Rev. Lett.}\ }\textbf
  {\bibinfo {volume} {85}},\ \bibinfo {pages} {4972} (\bibinfo {year}
  {2000})}\BibitemShut {NoStop}%
\bibitem [{\citenamefont {A.Wehrl}(1978)}]{rel_entropy}%
  \BibitemOpen
  \bibfield  {author} {\bibinfo {author} {\bibnamefont {A.Wehrl}},\ }\bibfield
  {title} {\bibinfo {title} {General properties of entropy},\ }\href
  {https://link.aps.org/doi/10.1103/RevModPhys.50.221} {\bibfield  {journal}
  {\bibinfo  {journal} {Rev. Mod. Phys.}\ }\textbf {\bibinfo {volume} {50}},\
  \bibinfo {pages} {221} (\bibinfo {year} {1978})}\BibitemShut {NoStop}%
\bibitem [{\citenamefont {V.Vedral}(2002)}]{vedral_ent}%
  \BibitemOpen
  \bibfield  {author} {\bibinfo {author} {\bibnamefont {V.Vedral}},\ }\bibfield
   {title} {\bibinfo {title} {The role of relative entropy in quantum
  information theory},\ }\href
  {https://link.aps.org/doi/10.1103/RevModPhys.74.197} {\bibfield  {journal}
  {\bibinfo  {journal} {Rev. Mod. Phys.}\ }\textbf {\bibinfo {volume} {74}},\
  \bibinfo {pages} {197} (\bibinfo {year} {2002})}\BibitemShut {NoStop}%
\bibitem [{\citenamefont {Wei}\ and\ \citenamefont {Goldbart}(2003)}]{GM}%
  \BibitemOpen
  \bibfield  {author} {\bibinfo {author} {\bibfnamefont {T.-C.}\ \bibnamefont
  {Wei}}\ and\ \bibinfo {author} {\bibfnamefont {P.~M.}\ \bibnamefont
  {Goldbart}},\ }\bibfield  {title} {\bibinfo {title} {Geometric measure of
  entanglement and applications to bipartite and multipartite quantum states},\
  }\href {https://link.aps.org/doi/10.1103/PhysRevA.68.042307} {\bibfield
  {journal} {\bibinfo  {journal} {Phys. Rev. A}\ }\textbf {\bibinfo {volume}
  {68}},\ \bibinfo {pages} {042307} (\bibinfo {year} {2003})}\BibitemShut
  {NoStop}%
\bibitem [{\citenamefont {Blasone}\ \emph {et~al.}(2008)\citenamefont
  {Blasone}, \citenamefont {Dell'Anno}, \citenamefont {De~Siena},\ and\
  \citenamefont {Illuminati}}]{Blasone2008}%
  \BibitemOpen
  \bibfield  {author} {\bibinfo {author} {\bibfnamefont {M.}~\bibnamefont
  {Blasone}}, \bibinfo {author} {\bibfnamefont {F.}~\bibnamefont {Dell'Anno}},
  \bibinfo {author} {\bibfnamefont {S.}~\bibnamefont {De~Siena}},\ and\
  \bibinfo {author} {\bibfnamefont {F.}~\bibnamefont {Illuminati}},\ }\bibfield
   {title} {\bibinfo {title} {Hierarchies of geometric entanglement},\ }\href
  {https://doi.org/10.1103/PhysRevA.77.062304} {\bibfield  {journal} {\bibinfo
  {journal} {Phys. Rev. A}\ }\textbf {\bibinfo {volume} {77}},\ \bibinfo
  {pages} {062304} (\bibinfo {year} {2008})}\BibitemShut {NoStop}%
\bibitem [{\citenamefont {Cianciaruso}\ \emph {et~al.}(2016)\citenamefont
  {Cianciaruso}, \citenamefont {Bromley},\ and\ \citenamefont
  {Adesso}}]{Cianciaruso2016}%
  \BibitemOpen
  \bibfield  {author} {\bibinfo {author} {\bibfnamefont {M.}~\bibnamefont
  {Cianciaruso}}, \bibinfo {author} {\bibfnamefont {T.~R.}\ \bibnamefont
  {Bromley}},\ and\ \bibinfo {author} {\bibfnamefont {G.}~\bibnamefont
  {Adesso}},\ }\bibfield  {title} {\bibinfo {title} {Accessible quantification
  of multiparticle entanglement},\ }\href
  {https://doi.org/10.1038/npjqi.2016.30} {\bibfield  {journal} {\bibinfo
  {journal} {npj Quantum Information}\ }\textbf {\bibinfo {volume} {2}},\
  \bibinfo {pages} {2056} (\bibinfo {year} {2016})}\BibitemShut {NoStop}%
\bibitem [{\citenamefont {Friis}\ \emph {et~al.}(2019)\citenamefont {Friis},
  \citenamefont {Vitagliano}, \citenamefont {Malik},\ and\ \citenamefont
  {Huber}}]{Friis2019}%
  \BibitemOpen
  \bibfield  {author} {\bibinfo {author} {\bibfnamefont {N.}~\bibnamefont
  {Friis}}, \bibinfo {author} {\bibfnamefont {G.}~\bibnamefont {Vitagliano}},
  \bibinfo {author} {\bibfnamefont {M.}~\bibnamefont {Malik}},\ and\ \bibinfo
  {author} {\bibfnamefont {M.}~\bibnamefont {Huber}},\ }\bibfield  {title}
  {\bibinfo {title} {Entanglement certification from theory to experiment},\
  }\href {https://doi.org/10.1038/s42254-018-0003-5} {\bibfield  {journal}
  {\bibinfo  {journal} {Nature Reviews Physics}\ }\textbf {\bibinfo {volume}
  {1}},\ \bibinfo {pages} {72} (\bibinfo {year} {2019})}\BibitemShut {NoStop}%
\bibitem [{\citenamefont {Greenberger}\ \emph {et~al.}(1989)\citenamefont
  {Greenberger}, \citenamefont {Horne},\ and\ \citenamefont
  {Zeilinger}}]{Greenberger1989}%
  \BibitemOpen
  \bibfield  {author} {\bibinfo {author} {\bibfnamefont {D.~M.}\ \bibnamefont
  {Greenberger}}, \bibinfo {author} {\bibfnamefont {M.~A.}\ \bibnamefont
  {Horne}},\ and\ \bibinfo {author} {\bibfnamefont {A.}~\bibnamefont
  {Zeilinger}},\ }\bibfield  {title} {\bibinfo {title} {Going beyond bell's
  theorem},\ }\href@noop {} {\bibfield  {journal} {\bibinfo  {journal} {in:
  'Bell's Theorem, Quantum Theory, and Conceptions of the Universe, Edited by
  M. Kafatos (Ed.), Kluwer, Dordrecht}\ ,\ \bibinfo {pages} {p. 69}} (\bibinfo
  {year} {1989})}\BibitemShut {NoStop}%
\bibitem [{\citenamefont {Mermin}(1990)}]{Mermin1990}%
  \BibitemOpen
  \bibfield  {author} {\bibinfo {author} {\bibfnamefont {N.~D.}\ \bibnamefont
  {Mermin}},\ }\bibfield  {title} {\bibinfo {title} {Quantum mysteries
  revisited},\ }\href {https://doi.org/10.1119/1.16503} {\bibfield  {journal}
  {\bibinfo  {journal} {Am. J. Phys.}\ }\textbf {\bibinfo {volume} {58}},\
  \bibinfo {pages} {731} (\bibinfo {year} {1990})}\BibitemShut {NoStop}%
\bibitem [{\citenamefont {H\"ubener}\ \emph {et~al.}(2009)\citenamefont
  {H\"ubener}, \citenamefont {Kleinmann}, \citenamefont {Wei}, \citenamefont
  {Gonz\'alez-Guill\'en},\ and\ \citenamefont {G\"uhne}}]{GMsymm}%
  \BibitemOpen
  \bibfield  {author} {\bibinfo {author} {\bibfnamefont {R.}~\bibnamefont
  {H\"ubener}}, \bibinfo {author} {\bibfnamefont {M.}~\bibnamefont
  {Kleinmann}}, \bibinfo {author} {\bibfnamefont {T.-C.}\ \bibnamefont {Wei}},
  \bibinfo {author} {\bibfnamefont {C.}~\bibnamefont {Gonz\'alez-Guill\'en}},\
  and\ \bibinfo {author} {\bibfnamefont {O.}~\bibnamefont {G\"uhne}},\
  }\bibfield  {title} {\bibinfo {title} {Geometric measure of entanglement for
  symmetric states},\ }\href {https://doi.org/10.1103/PhysRevA.80.032324}
  {\bibfield  {journal} {\bibinfo  {journal} {Phys. Rev. A}\ }\textbf {\bibinfo
  {volume} {80}},\ \bibinfo {pages} {032324} (\bibinfo {year}
  {2009})}\BibitemShut {NoStop}%
\bibitem [{\citenamefont {Chen}\ \emph {et~al.}(2010)\citenamefont {Chen},
  \citenamefont {Xu},\ and\ \citenamefont {Zhu}}]{zhu2010}%
  \BibitemOpen
  \bibfield  {author} {\bibinfo {author} {\bibfnamefont {L.}~\bibnamefont
  {Chen}}, \bibinfo {author} {\bibfnamefont {A.}~\bibnamefont {Xu}},\ and\
  \bibinfo {author} {\bibfnamefont {H.}~\bibnamefont {Zhu}},\ }\bibfield
  {title} {\bibinfo {title} {Computation of the geometric measure of
  entanglement for pure multiqubit states},\ }\href
  {https://doi.org/10.1103/PhysRevA.82.032301} {\bibfield  {journal} {\bibinfo
  {journal} {Phys. Rev. A}\ }\textbf {\bibinfo {volume} {82}},\ \bibinfo
  {pages} {032301} (\bibinfo {year} {2010})}\BibitemShut {NoStop}%
\bibitem [{\citenamefont {Hu}\ \emph {et~al.}(2016)\citenamefont {Hu},
  \citenamefont {Qi},\ and\ \citenamefont {Zhang}}]{Zhang2016}%
  \BibitemOpen
  \bibfield  {author} {\bibinfo {author} {\bibfnamefont {S.}~\bibnamefont
  {Hu}}, \bibinfo {author} {\bibfnamefont {L.}~\bibnamefont {Qi}},\ and\
  \bibinfo {author} {\bibfnamefont {G.}~\bibnamefont {Zhang}},\ }\bibfield
  {title} {\bibinfo {title} {Computing the geometric measure of entanglement of
  multipartite pure states by means of non-negative tensors},\ }\href
  {https://doi.org/10.1103/PhysRevA.93.012304} {\bibfield  {journal} {\bibinfo
  {journal} {Phys. Rev. A}\ }\textbf {\bibinfo {volume} {93}},\ \bibinfo
  {pages} {012304} (\bibinfo {year} {2016})}\BibitemShut {NoStop}%
\bibitem [{\citenamefont {Sen(De)}\ and\ \citenamefont
  {Sen}(2010)}]{Sen(De)2010}%
  \BibitemOpen
  \bibfield  {author} {\bibinfo {author} {\bibfnamefont {A.}~\bibnamefont
  {Sen(De)}}\ and\ \bibinfo {author} {\bibfnamefont {U.}~\bibnamefont {Sen}},\
  }\bibfield  {title} {\bibinfo {title} {Channel capacities versus entanglement
  measures in multiparty quantum states},\ }\href
  {https://doi.org/10.1103/PhysRevA.81.012308} {\bibfield  {journal} {\bibinfo
  {journal} {Phys. Rev. A}\ }\textbf {\bibinfo {volume} {81}},\ \bibinfo
  {pages} {012308} (\bibinfo {year} {2010})}\BibitemShut {NoStop}%
\bibitem [{\citenamefont {Sen(De)}\ and\ \citenamefont {Sen}()}]{Sen2010}%
  \BibitemOpen
  \bibfield  {author} {\bibinfo {author} {\bibfnamefont {A.}~\bibnamefont
  {Sen(De)}}\ and\ \bibinfo {author} {\bibfnamefont {U.}~\bibnamefont {Sen}},\
  }\bibfield  {title} {\bibinfo {title} {Bound genuine multisite entanglement:
  Detector of gapless-gapped quantum transitions in frustrated systems},\
  }\href@noop {} {\bibinfo  {journal} {arXiv:1002.1253}\ }\BibitemShut
  {NoStop}%
\bibitem [{\citenamefont {Biswas}\ \emph {et~al.}(2014)\citenamefont {Biswas},
  \citenamefont {Prabhu}, \citenamefont {Sen(De)},\ and\ \citenamefont
  {Sen}}]{Biswas2014}%
  \BibitemOpen
\bibfield  {journal} {  }\bibfield  {author} {\bibinfo {author} {\bibfnamefont
  {A.}~\bibnamefont {Biswas}}, \bibinfo {author} {\bibfnamefont
  {R.}~\bibnamefont {Prabhu}}, \bibinfo {author} {\bibfnamefont
  {A.}~\bibnamefont {Sen(De)}},\ and\ \bibinfo {author} {\bibfnamefont
  {U.}~\bibnamefont {Sen}},\ }\bibfield  {title} {\bibinfo {title}
  {Genuine-multipartite-entanglement trends in gapless-to-gapped transitions of
  quantum spin systems},\ }\href {https://doi.org/10.1103/PhysRevA.90.032301}
  {\bibfield  {journal} {\bibinfo  {journal} {Phys. Rev. A}\ }\textbf {\bibinfo
  {volume} {90}},\ \bibinfo {pages} {032301} (\bibinfo {year}
  {2014})}\BibitemShut {NoStop}%
\bibitem [{\citenamefont {Das}\ \emph {et~al.}(2016)\citenamefont {Das},
  \citenamefont {Roy}, \citenamefont {Bagchi}, \citenamefont {Misra},
  \citenamefont {Sen(De)},\ and\ \citenamefont {Sen}}]{GGM}%
  \BibitemOpen
  \bibfield  {author} {\bibinfo {author} {\bibfnamefont {T.}~\bibnamefont
  {Das}}, \bibinfo {author} {\bibfnamefont {S.~S.}\ \bibnamefont {Roy}},
  \bibinfo {author} {\bibfnamefont {S.}~\bibnamefont {Bagchi}}, \bibinfo
  {author} {\bibfnamefont {A.}~\bibnamefont {Misra}}, \bibinfo {author}
  {\bibfnamefont {A.}~\bibnamefont {Sen(De)}},\ and\ \bibinfo {author}
  {\bibfnamefont {U.}~\bibnamefont {Sen}},\ }\bibfield  {title} {\bibinfo
  {title} {Generalized geometric measure of entanglement for multiparty mixed
  states},\ }\href {https://link.aps.org/doi/10.1103/PhysRevA.94.022336}
  {\bibfield  {journal} {\bibinfo  {journal} {Phys. Rev. A}\ }\textbf {\bibinfo
  {volume} {94}},\ \bibinfo {pages} {022336} (\bibinfo {year}
  {2016})}\BibitemShut {NoStop}%
\bibitem [{\citenamefont {{Liu}}\ \emph {et~al.}(2017)\citenamefont {{Liu}},
  \citenamefont {{Zhang}}, \citenamefont {{Yu}}, \citenamefont {{Ding}},\ and\
  \citenamefont {{Liu}}}]{fid}%
  \BibitemOpen
  \bibfield  {author} {\bibinfo {author} {\bibfnamefont {C.~L.}\ \bibnamefont
  {{Liu}}}, \bibinfo {author} {\bibfnamefont {D.-J.}\ \bibnamefont {{Zhang}}},
  \bibinfo {author} {\bibfnamefont {X.-D.}\ \bibnamefont {{Yu}}}, \bibinfo
  {author} {\bibfnamefont {Q.-M.}\ \bibnamefont {{Ding}}},\ and\ \bibinfo
  {author} {\bibfnamefont {L.}~\bibnamefont {{Liu}}},\ }\bibfield  {title}
  {\bibinfo {title} {{A new coherence measure based on fidelity}},\ }\href
  {https://doi.org/10.1007/s11128-017-1650-7} {\bibfield  {journal} {\bibinfo
  {journal} {Quantum Information Processing}\ }\textbf {\bibinfo {volume}
  {16}},\ \bibinfo {eid} {198} (\bibinfo {year} {2017})}\BibitemShut {NoStop}%
\bibitem [{\citenamefont {Moreno}\ and\ \citenamefont
  {Parisio}(2017)}]{parisio2017}%
  \BibitemOpen
  \bibfield  {author} {\bibinfo {author} {\bibfnamefont {M.~G.~M.}\
  \bibnamefont {Moreno}}\ and\ \bibinfo {author} {\bibfnamefont
  {F.}~\bibnamefont {Parisio}},\ }\bibfield  {title} {\bibinfo {title}
  {Critical behaviour in the optimal generation of multipartite entanglement},\
  }\href {https://doi.org/10.1038/s41598-017-06299-5} {\bibfield  {journal}
  {\bibinfo  {journal} {Scientific Reports}\ }\textbf {\bibinfo {volume} {7}},\
  \bibinfo {pages} {6259} (\bibinfo {year} {2017})}\BibitemShut {NoStop}%
\bibitem [{\citenamefont {Yamasaki}\ \emph {et~al.}(2018)\citenamefont
  {Yamasaki}, \citenamefont {Pirker}, \citenamefont {Murao}, \citenamefont
  {D\"ur},\ and\ \citenamefont {Kraus}}]{kraus2018}%
  \BibitemOpen
  \bibfield  {author} {\bibinfo {author} {\bibfnamefont {H.}~\bibnamefont
  {Yamasaki}}, \bibinfo {author} {\bibfnamefont {A.}~\bibnamefont {Pirker}},
  \bibinfo {author} {\bibfnamefont {M.}~\bibnamefont {Murao}}, \bibinfo
  {author} {\bibfnamefont {W.}~\bibnamefont {D\"ur}},\ and\ \bibinfo {author}
  {\bibfnamefont {B.}~\bibnamefont {Kraus}},\ }\bibfield  {title} {\bibinfo
  {title} {Multipartite entanglement outperforming bipartite entanglement under
  limited quantum system sizes},\ }\href
  {https://doi.org/10.1103/PhysRevA.98.052313} {\bibfield  {journal} {\bibinfo
  {journal} {Phys. Rev. A}\ }\textbf {\bibinfo {volume} {98}},\ \bibinfo
  {pages} {052313} (\bibinfo {year} {2018})}\BibitemShut {NoStop}%
\bibitem [{\citenamefont {Nezami}\ and\ \citenamefont {Walter}(2020)}]{multi}%
  \BibitemOpen
  \bibfield  {author} {\bibinfo {author} {\bibfnamefont {S.}~\bibnamefont
  {Nezami}}\ and\ \bibinfo {author} {\bibfnamefont {M.}~\bibnamefont
  {Walter}},\ }\bibfield  {title} {\bibinfo {title} {Multipartite entanglement
  in stabilizer tensor networks},\ }\href
  {https://doi.org/10.1103/PhysRevLett.125.241602} {\bibfield  {journal}
  {\bibinfo  {journal} {Phys. Rev. Lett.}\ }\textbf {\bibinfo {volume} {125}},\
  \bibinfo {pages} {241602} (\bibinfo {year} {2020})}\BibitemShut {NoStop}%
\bibitem [{\citenamefont {Halder}\ \emph {et~al.}(2019)\citenamefont {Halder},
  \citenamefont {Banik}, \citenamefont {Agrawal},\ and\ \citenamefont
  {Bandyopadhyay}}]{multiloc}%
  \BibitemOpen
  \bibfield  {author} {\bibinfo {author} {\bibfnamefont {S.}~\bibnamefont
  {Halder}}, \bibinfo {author} {\bibfnamefont {M.}~\bibnamefont {Banik}},
  \bibinfo {author} {\bibfnamefont {S.}~\bibnamefont {Agrawal}},\ and\ \bibinfo
  {author} {\bibfnamefont {S.}~\bibnamefont {Bandyopadhyay}},\ }\bibfield
  {title} {\bibinfo {title} {Strong quantum nonlocality without entanglement},\
  }\href {https://doi.org/10.1103/PhysRevLett.122.040403} {\bibfield  {journal}
  {\bibinfo  {journal} {Phys. Rev. Lett.}\ }\textbf {\bibinfo {volume} {122}},\
  \bibinfo {pages} {040403} (\bibinfo {year} {2019})}\BibitemShut {NoStop}%
\bibitem [{\citenamefont {Napoli}\ \emph {et~al.}(2016)\citenamefont {Napoli},
  \citenamefont {Bromley}, \citenamefont {Cianciaruso}, \citenamefont {Piani},
  \citenamefont {Johnston},\ and\ \citenamefont {Adesso}}]{Napoli2016}%
  \BibitemOpen
  \bibfield  {author} {\bibinfo {author} {\bibfnamefont {C.}~\bibnamefont
  {Napoli}}, \bibinfo {author} {\bibfnamefont {T.~R.}\ \bibnamefont {Bromley}},
  \bibinfo {author} {\bibfnamefont {M.}~\bibnamefont {Cianciaruso}}, \bibinfo
  {author} {\bibfnamefont {M.}~\bibnamefont {Piani}}, \bibinfo {author}
  {\bibfnamefont {N.}~\bibnamefont {Johnston}},\ and\ \bibinfo {author}
  {\bibfnamefont {G.}~\bibnamefont {Adesso}},\ }\bibfield  {title} {\bibinfo
  {title} {Robustness of coherence: An operational and observable measure of
  quantum coherence},\ }\href {https://doi.org/10.1103/PhysRevLett.116.150502}
  {\bibfield  {journal} {\bibinfo  {journal} {Phys. Rev. Lett.}\ }\textbf
  {\bibinfo {volume} {116}},\ \bibinfo {pages} {150502} (\bibinfo {year}
  {2016})}\BibitemShut {NoStop}%
\bibitem [{\citenamefont {Cimini}\ \emph {et~al.}(2019)\citenamefont {Cimini},
  \citenamefont {Gianani}, \citenamefont {Sbroscia}, \citenamefont {Sperling},\
  and\ \citenamefont {Barbieri}}]{Cimini2019}%
  \BibitemOpen
  \bibfield  {author} {\bibinfo {author} {\bibfnamefont {V.}~\bibnamefont
  {Cimini}}, \bibinfo {author} {\bibfnamefont {I.}~\bibnamefont {Gianani}},
  \bibinfo {author} {\bibfnamefont {M.}~\bibnamefont {Sbroscia}}, \bibinfo
  {author} {\bibfnamefont {J.}~\bibnamefont {Sperling}},\ and\ \bibinfo
  {author} {\bibfnamefont {M.}~\bibnamefont {Barbieri}},\ }\bibfield  {title}
  {\bibinfo {title} {Measuring coherence of quantum measurements},\ }\href
  {https://doi.org/10.1103/PhysRevResearch.1.033020} {\bibfield  {journal}
  {\bibinfo  {journal} {Phys. Rev. Res.}\ }\textbf {\bibinfo {volume} {1}},\
  \bibinfo {pages} {033020} (\bibinfo {year} {2019})}\BibitemShut {NoStop}%
\bibitem [{\citenamefont {Horodecki}\ \emph
  {et~al.}(2003{\natexlab{b}})\citenamefont {Horodecki}, \citenamefont
  {Horodecki},\ and\ \citenamefont {Oppenheim}}]{HHO2003}%
  \BibitemOpen
  \bibfield  {author} {\bibinfo {author} {\bibfnamefont {M.}~\bibnamefont
  {Horodecki}}, \bibinfo {author} {\bibfnamefont {P.}~\bibnamefont
  {Horodecki}},\ and\ \bibinfo {author} {\bibfnamefont {J.}~\bibnamefont
  {Oppenheim}},\ }\bibfield  {title} {\bibinfo {title} {Reversible
  transformations from pure to mixed states and the unique measure of
  information},\ }\href {https://doi.org/10.1103/PhysRevA.67.062104} {\bibfield
   {journal} {\bibinfo  {journal} {Phys. Rev. A}\ }\textbf {\bibinfo {volume}
  {67}},\ \bibinfo {pages} {062104} (\bibinfo {year}
  {2003}{\natexlab{b}})}\BibitemShut {NoStop}%
\end{thebibliography}%

\end{document}